\documentclass[sigconf]{acmart}
\usepackage{graphicx} %
\usepackage{xcolor}
\usepackage{lipsum}
\usepackage{array}
\usepackage{multirow}
\usepackage{booktabs}
\usepackage{subcaption}
\usepackage{algorithm}
\usepackage{natbib}
\usepackage{algpseudocode}
\usepackage{makecell}
\usepackage{tikz} 
\usetikzlibrary{arrows.meta}
\usepackage{xspace}
\usepackage{textpos}
\usepackage{url}
\usepackage{colortbl} %
\usepackage{array}      %
\usepackage{amssymb}
\usepackage{caption} %
\usepackage{xfrac}

\newcommand{\cmark}{\textcolor{green}{\checkmark}}
\newcommand{\xmark}{\textcolor{red}{\textsf{X}}}



\makeatletter
\renewcommand*{\@fnsymbol}[1]{%
  \ensuremath{%
    \ifcase#1\or \dagger\or \ast\or \ddagger\or
    \mathsection\or \mathparagraph\or \|\or **\or \dagger\dagger
    \or \ddagger\ddagger \else\@ctrerr\fi}}
\makeatother

\newcommand{\acq}{\alpha}
\newcommand{\ei}{\mathrm{EI}}

\newcommand{\design}{\mathbf{x}}

\newcommand{\name}{\textsc{Cabop}\xspace}

\definecolor{mygreen}{HTML}{82BD76}  
\definecolor{myorange}{HTML}{EABF49}  
\definecolor{myred}{HTML}{E63D3D}     

\newcommand{\tweak}{\textcolor{mygreen}{\textbf{tweak}}\xspace}
\newcommand{\swap}{\textcolor{myorange}{\textbf{swap}}\xspace}
\newcommand{\create}{\textcolor{myred}{\textbf{create}}\xspace}

\newcommand{\Sec}[1]{Sec.~\ref{#1}}
\newcommand{\Eq}[1]{Eq.~\ref{#1}}
\newcommand{\Fig}[1]{Fig.~\ref{#1}}
\newcommand{\App}[1]{Appendix~\ref{#1}}
\newcommand{\Tab}[1]{Tab.~\ref{#1}}
\newcommand{\Alg}[1]{Alg.~\ref{#1}}

\usepackage{subcaption}
\usepackage{tabularx}

\newif\ifshowadd
\newif\ifshowcr

\newcommand{\add}[1]{%
  \ifshowadd
    {\color{blue}#1}%
  \else
    #1%
  \fi
}

\usepackage[normalem]{ulem}

\newcommand{\addcr}[1]{%
  \ifshowcr
    {\color{blue}#1}%
  \else
    #1%
  \fi
}

\showaddfalse
\showcrfalse

\begin{document}

\title{Cost-Aware Bayesian Optimization for Prototyping Interactive Devices}

\author{Thomas Langerak}
\orcid{0000-0003-2536-0208}
\affiliation{%
  \institution{Aalto University}
  \city{Helsinki}
  \country{Finland}
}
\author{Renate Zhang}
\orcid{0009-0002-5291-3103}
\authornote{Work done while interning at Aalto University. Also with TU Wien, Vienna, Austria.}
\affiliation{%
  \institution{Aalto University}
  \city{Helsinki}
  \country{Finland}
}

\author{Ziyuan Wang}
\orcid{0009-0008-1534-8487}
\affiliation{%
  \institution{Aalto University}
  \city{Helsinki}
  \country{Finland}
}

\author{Per Ola Kristensson}
\orcid{0000-0002-7139-871X}
\affiliation{%
  \institution{University of Cambridge}
  \city{Cambridge}
  \country{United Kingdom}
}

\author{Antti Oulasvirta}
\orcid{0000-0002-2498-7837}
\affiliation{%
  \institution{Aalto University \& ELLIS Institute}
  \city{Helsinki}
  \country{Finland}
}

\copyrightyear{2026}
\acmYear{2026}
\setcopyright{cc}
\setcctype{by}
\acmConference[CHI '26]{Proceedings of the 2026 CHI Conference on Human Factors in Computing Systems}{April 13--17, 2026}{Barcelona, Spain}
\acmBooktitle{Proceedings of the 2026 CHI Conference on Human Factors in Computing Systems (CHI '26), April 13--17, 2026, Barcelona, Spain}
\acmDOI{10.1145/3772318.3791024}
\acmISBN{979-8-4007-2278-3/2026/04}

\begin{teaserfigure}
  \centering
  \includegraphics[width=0.9\linewidth]{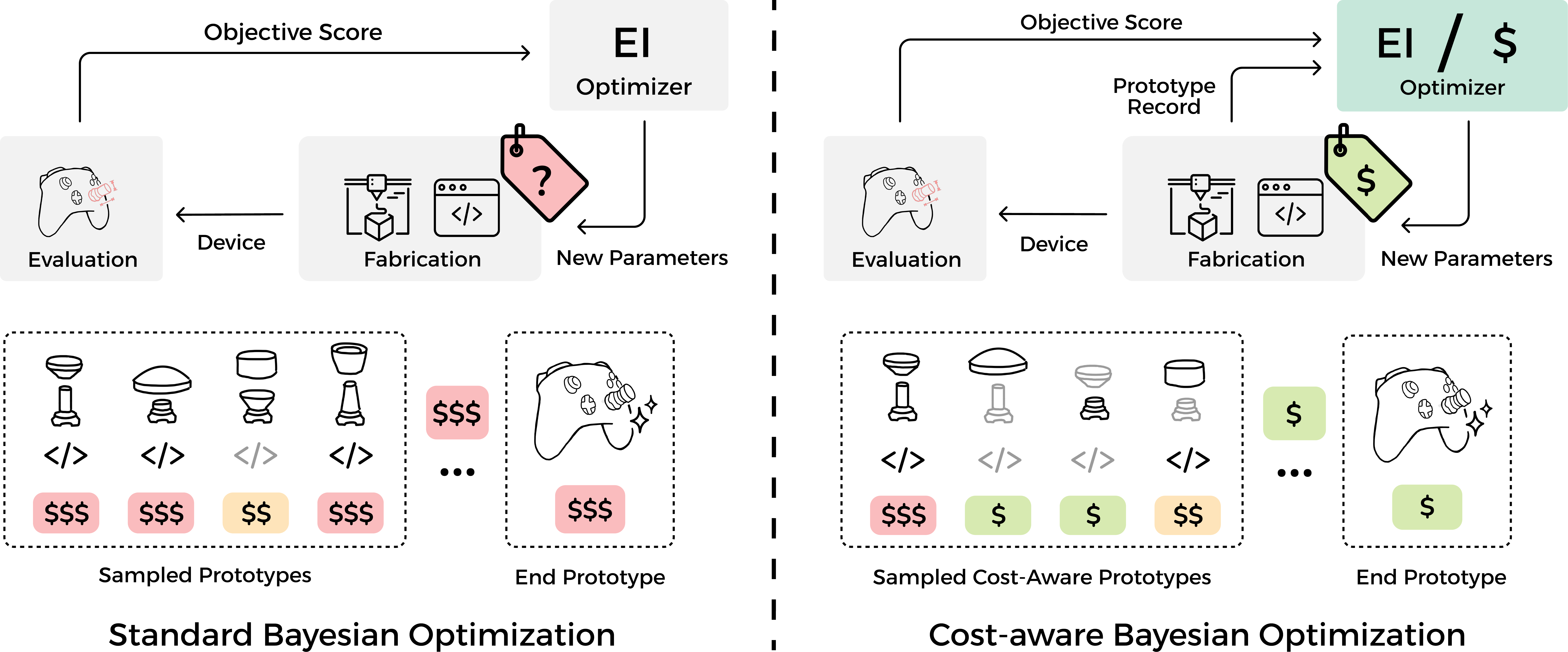}
    \vspace{0.5em}
    \caption{We extend cost-aware Bayesian optimization method for prototyping interactive devices. It accounts for the different costs of hardware and software design, while reusing existing prototypes stored in a record. By embedding a cost model into the acquisition function, the optimizer can balance Expected Improvement (EI) against cost. Because our approach operates only at the acquisition function level, it does not influence the modeling process and surrogate model, making it broadly applicable.
    }
    \Description{Comparison of standard versus cost-aware Bayesian optimization.
  The left panel illustrates standard Bayesian optimization, where the optimizer selects new parameters based on expected improvement, leading to repeated fabrication and evaluation of prototypes with varying costs. This results in many expensive sampled prototypes before reaching the final design. The right panel shows cost-aware Bayesian optimization, where the optimizer considers expected improvement per unit cost. By incorporating prototype records and cost information, it favors reusing or selecting lower-cost prototypes, resulting in more cost-efficient sampling and a cheaper final prototype.}
\label{fig:teaser}
\end{teaserfigure}

\begin{abstract}
Deciding which idea is worth prototyping is a central concern in iterative design. 
A prototype should be produced when the expected improvement is high and the cost is low. 
However, this is hard to decide, because costs can vary drastically: a simple parameter tweak may take seconds, while fabricating hardware \addcr{consumes material and energy}. Such asymmetries, can discourage a designer from exploring the design space.
In this paper, we present \addcr{an extension} of cost-aware Bayesian optimization to account for diverse prototyping costs. 
The method builds on the power of Bayesian optimization and requires only a minimal modification to the acquisition function.
The key idea is to use designer-estimated costs to guide sampling toward more cost-effective prototypes. 
In technical evaluations, the method achieved comparable utility to a cost-agnostic baseline while requiring only ${\approx}70\%$ of the cost; under strict budgets, it outperformed the baseline threefold. A within-subjects study with 12 participants in a realistic joystick design task demonstrated similar benefits. These results show that accounting for prototyping costs can make Bayesian optimization more compatible with real-world design projects.

\end{abstract}

\maketitle

\section{Introduction}
\enlargethispage{2\baselineskip}
Cost is not just one factor among many in prototyping; it decisively shapes what gets prototyped and, in turn, strongly influences the final outcome.
When building a prototype, \emph{everything carries a price tag}: time, labor, materials, and external services.
Take a game controller, for example: costs accumulate across fabrication, hardware acquisition, electronics design, software development, and integration engineering. Depending on the level of fidelity, the bill may run anywhere from a few hundred dollars to tens of thousands.
Hardware costs, in particular, are large, and surface as a central concern across HCI research domains such as fabrication \cite{virzi1989can}, ubiquitous computing \cite{weiser1998future}, tangible computing \cite{holmquist2023bits}, assistive technology \cite{hofmann2016clinical}, DIY communities \cite{lindtner2014emerging}, and interactive devices \cite{langerak2020omni, langerak2020optimal}.
Making physical things---enclosures, actuators, sensors, and displays---demands material, time, and energy. 
As Holmquist laments, \emph{''bits are cheap, atoms are expensive''} \cite{holmquist2023bits}.
Such costs are consequential. They constrain not only which prototypes are explored, but also how many iterations a team attempts, increasing pressure to get it right as cheaply \addcr{and sustainably} as possible.

In this paper, we examine a direction that is complementary to existing work.
Efforts in HCI to reduce prototyping costs generally fall into three categories: toolkits that make iteration more efficient by abstracting design choices \cite{alexander2018grand}; fabrication techniques that cut down material or time costs \cite{tao2017weavemesh}; and computational approaches that estimate the feasibility or performance of a design candidate before it is built \cite{chan2022investigating, feit2021}.
Collectively, these approaches aim to reduce the cost of implementing a given prototype idea. What they often leave unexamined is how cost influences which ideas are considered worth prototyping at all.

We \addcr{study and extend a} cost-aware method specifically for the exploration--exploitation problem in prototyping.
When prototyping, a designer must constantly decide whether to try something novel (exploration) or to refine a candidate that already looks promising (exploitation).
Bayesian optimization has recently gained traction in HCI research as a principled approach to balancing these choices \cite{liao2023interaction, khajah2016designing, dudley2019crowdsourcing, koyama2022bo}. 
It is sample-efficient, considers uncertainty, supports multiple objectives, and allows different techniques for steering it \cite{frazier2018bayesian}. 
While working with an optimizer can demonstrably help designers better explore a design space \cite{liao2023human}, existing work ignores costs.
In scenarios where every prototype is roughly equal in expense, this simplification is acceptable.
However, in real-world projects, where costs are heterogeneous and highly variable, a 'cost-blind' method can be useless.
It overlooks cases where producing a single expensive prototype could exhaust the budget far more severely than simply tweaking an existing design.
We believe that the critical decision---what is worth trying next---should be approached as a resource allocation problem.

\begin{figure*}
\centering
\includegraphics[width=0.85\linewidth]{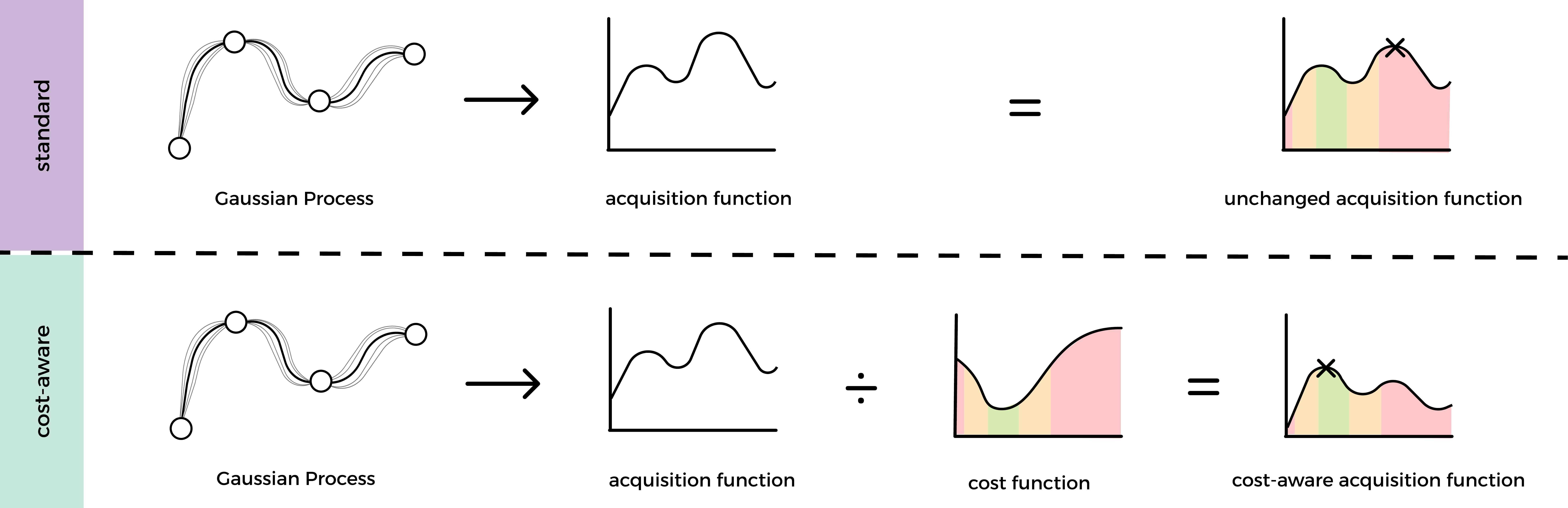}
\vspace{1.0em}
\caption{
\textbf{Top:} In standard Bayesian optimization, a surrogate model (e.g., a Gaussian Process) predicts utility, and the \emph{expected improvement} (EI) directly defines the acquisition function. This approach is \emph{cost-blind}; it treats all evaluations as equally expensive.
\textbf{Bottom:} Cost-aware Bayesian optimization incorporates a cost model by dividing EI by the estimated cost $c(\design)$, yielding \emph{expected improvement per unit cost} \addcr{\cite{lee2020cost}}. This shifts the focus from being purely sample-efficient to being cost-efficient, prioritizing low-cost, high-improvement evaluations. \addcr{Colors correspond to a \tweak (\$), \swap (\$\$), or \create (\$\$\$) parameter}.
}
\Description{Standard versus cost-aware acquisition functions in Bayesian optimization.
The top row shows the standard approach: a Gaussian Process generates an acquisition function, which remains unchanged and directly guides sampling. The bottom row shows the cost-aware variant: the acquisition function is divided by a cost function, producing a cost-aware acquisition function that balances improvement against cost, illustrated by colored regions and a shifted sampling point.
}
\label{fig:deltabo}
\end{figure*}

\name is a \underline{c}ost-\underline{a}ware extension of \underline{B}ayesian \underline{o}ptimization for \underline{p}rototyping interactive devices (\Fig{fig:teaser}).
It selects the next design by maximizing expected improvement per unit cost, enabling more efficient exploration under realistic budget considerations. \add{As with existing HCI applications of Bayesian optimization (e.g.,\cite{koyama2022bo,liao2023interaction, dudley2019crowdsourcing}), our method assumes a parameterized design space provided in advance.}
Our main contribution \addcr{is adapting and contextualizing the} \emph{acquisition function} for efficient prototyping, illustrated in \Fig{fig:deltabo}.
The acquisition function is the ‘heart’ of Bayesian optimization: it decides which sample to try next given what has been learned so far.
In standard Bayesian optimization, the acquisition function selects the next design by maximizing expected utility; typically, expected improvement in performance.
\addcr{Cost-aware Bayesian optimization \cite{lee2020cost} modifies this objective to maximize} for \emph{expected improvement per unit cost}:
\begin{equation}
\acq(\design) = \frac{\ei(\design)}{c(\design)},
\end{equation}
where $\ei(\design)$ is the expected improvement of design $\design$ and $c(\design)$ is the estimated cost.
While this formulation has precedents in the optimization literature, our contribution is to adapt and operationalize it for HCI prototyping.

We make two technical contributions to extend cost-awareness for prototyping.
First, while prototyping costs are highly varied, designers can often estimate them in advance. To enable the acquisition function to take advantage of this, we introduce a \emph{cost model}. It allows flexible modeling of the main prototyping costs in HCI projects. Each parameter in the design space can be assigned a cost.
For example, adjusting the gain function of a joystick might be relatively cheap (e.g., \$10), while altering the joystick's buttons or shell could require 3D printing and thus be an order of magnitude more expensive (e.g., \$100).
While the model allows arbitrary cost assignments, to guide designers we classify costs into three categories by introducing a \emph{cost classification}: \tweak (low-cost adjustments), \swap (reuse of existing subdesigns), and \create (producing a new solution).
Second, our model incorporates the prototypes that have already been produced: what we call the \emph{prototype record}. In many cases, past designs can be exploited to reduce future costs, whether through fabricated components, implemented software modules, or tested templates. These artifacts represent sunk costs that make subsequent iterations cheaper. From the perspective of the acquisition function, cost is therefore not fixed but depends on the prototype record; the set of prototypes already available that have been implemented and build, so that they can be reused.
Although this model does not capture interactions between components that may affect cost, it provides a practical way to account for the primary sources of expense. 

We evaluate our method both through simulations and in a user study with realistic hardware and software costs.
The results show that, compared to standard Bayesian optimization, our approach can substantially reduce costs without compromising outcome quality.
Under budget constraints, it identifies better-performing solutions than the standard method and similar performing solutions for ${\approx}70\%$ of the cumulative cost. Furthermore, we find that our approach dynamically adapts to a changing cost landscape, and prioritizes cheaper alternatives when possible. In the user study, participants achieved designs of comparable performance \add{for ${\approx}67\%$ of the original cost.} 

In summary, we contribute:
\begin{enumerate}
\item \addcr{The adaptation of} cost-aware Bayesian optimization to the domain of interactive devices, demonstrating how to navigate design spaces with diverse costs. 
\item A cost model for HCI projects that captures differences between software and hardware production while accounting for prototypes that can be reused.
\item Results from simulations and a user study that systematically compare the trade-offs of cost-aware optimization with the standard approach.
\end{enumerate}

\section{Related Work}
In HCI, cost is often the deciding factor in whether something is built at all. It influences whether a prototype is worth fabricating~\cite{virzi1989can}, whether a design is feasible in resource-constrained settings like DIY assistive technology~\cite{hofmann2016clinical}, or whether a field such as ubiquitous computing can scale~\cite{holmquist2023bits}. Beyond individual decisions, cost determines who can meaningfully participate in research and development, especially in under-resourced contexts~\cite{dell2016ins}, or whether there is even an opportunity for education in public settings like hackerspaces~\cite{lindtner2014emerging}. These costs are not theoretical. They directly determine what gets made, who gets to make it, and which ideas are explored or abandoned.

\subsection{Cost Reduction}
Prior work in HCI has approached the problem of prototyping cost from three main angles. First, toolkit research focuses on improving iteration efficiency by abstracting or simplifying design decisions~\cite{alexander2018grand, greenberg2007toolkits}. Toolkits generally target specific input or output modalities (e.g., low-cost eye tracking~\cite{winfield2005towards}) or are designed for specific audiences such as students~\cite{klemmer2009toolkit}. Alternatively, advances in fabrication methods aim to reduce material usage and production time, for example through more efficient physical construction techniques~\cite{tao2017weavemesh, abler2021hedgehog}, often tailored to contexts like makerspaces~\cite{camburn2018principles}. Finally, computational tools assist designers by predicting the feasibility or performance of candidate designs before committing to fabrication~\cite{chan2022investigating, feit2021}. Collectively, these efforts concentrate on lowering the cost of realizing a specific prototype configuration. Our approach complements this work by focusing not on making individual builds cheaper, but on improving the efficiency of design iteration. Rather than reducing the cost of making, we address the decision of \textit{what to make next} under real-world budget constraints.

\subsection{Bayesian Optimization in HCI}
Designers often balance exploration and exploitation when iterating on prototypes. Early sketches and low-fidelity mockups serve to explore cheaply and broadly, while higher-effort builds are reserved for refining promising directions. This tradeoff mirrors the logic of Bayesian optimization, which uses a probabilistic model to guide sample-efficient exploration of design spaces~\cite{frazier2018bayesian}. It has been applied in HCI to optimize interactive systems where evaluations are costly, noisy, or subjective. The approach uses a surrogate model to estimate performance and an acquisition function to select the next configuration, balancing uncertainty with performance.

Applications include animation~\cite{brochu2010bayesian}, wearable device design~\cite{kim2017human}, visual aesthetics~\cite{koyama2017sequential, koyama2020sequential}, photographic enhancement~\cite{yamamoto2022photographic}, mid-air gestures~\cite{yamamoto2022photographic}, typing~\cite{liao2026HOMI}, and typography~\cite{tatsukawa2025fontcraft}. These systems benefit from Bayesian optimization’s ability to search large spaces using few evaluations. ``Bayesian Optimization as Assistant'' further explores how this technique can support creative workflows where designers retain control over decision-making~\cite{koyama2022bo}. 

While these examples demonstrate the effectiveness of Bayesian optimization for guiding design iteration, most prior work do not take cost into account. In contrast, prototyping interactive systems introduces significant cost variation. Some changes may involve quick software edits, while others require fabrication time, material resources, or the reuse of existing components. These asymmetric and context-dependent costs are common in HCI practice but are not addressed by standard Bayesian optimization methods.

\add{Related extensions to Bayesian optimization address complex evaluation landscapes but do not directly solve this challenge. For instance, multi-objective optimization~\cite{jansen2025opti} can frame cost as an additional metric, yet this approach assumes cost is an unknown objective that must be modeled jointly with performance. In reality, designers often know or can estimate cost; a fact we exploit in this work. Similarly, multi-fidelity optimization~\cite{kandasamy2017multi} leverages evaluations of varying accuracy and expense, but it focuses on learning the bias between lower-fidelity approximations and the ground truth. In contrast, the prototyping costs described here are typically known and distinct from fidelity; they require an optimization strategy that allocates evaluations based on explicit, asymmetric resource costs rather than modeling unknown objectives or fidelity correlations.}

\subsection{Cost-Aware Bayesian Optimization}
Cost-aware Bayesian optimization has been explored outside of HCI to extend the standard formulation by incorporating cost directly into the acquisition function~\cite{hoffman2014correlation, lam2016bayesian, li2019bayesian}. Rather than selecting the configuration with the highest expected improvement, cost-aware variants aim to maximize the expected improvement per unit cost, enabling more efficient use of limited budgets and better prioritization of low-cost, high-value evaluations.
This approach has been applied across a variety of domains where costs are heterogeneous. In robotics, it reduces reliance on costly physical trials by combining them with simulated evaluations~\cite{yang2022bayesian}, and helps avoid unsafe configurations~\cite{berkenkamp2023bayesian}. In chemistry, it reduces reagent consumption and setup time~\cite{schoepfer2024cost}. In environmental monitoring, acquisition functions are adapted to minimize path length or energy usage~\cite{marchant2012bayesian}. In machine learning, it improves hyperparameter tuning by accounting for differences in training time and dataset size~\cite{snoek2012practical}.

However, these methods typically assume that cost is a fixed function of the input parameters. This assumption does not hold in interactive system design, where cost depends on context, such as the current configuration, component reuse, and prototype record. For example, a hardware parameter change may be cheap if a suitable component has already been fabricated, but expensive if it requires starting from scratch. These dynamic and prototype record costs are central to HCI prototyping workflows but are not modeled by conventional cost-aware Bayesian optimization techniques.
Our work addresses this gap by introducing a modular cost model that captures heterogeneous, context-sensitive costs. \add{While the $EI/C$ formulation itself follows established cost-aware optimization, what is specific to our setting is the structure of the cost term and its dependence on prototype reuse and design context.} We integrate this cost model into the Bayesian optimization process to support efficient exploration in iterative prototyping scenarios where cost is variable, asymmetric, and consequential.

\addcr{\section{Background: Cost-Aware Bayesian Optimization}

Bayesian optimization is a widely used method for the global optimization of black-box functions \cite{frazier2018bayesian}.
It is particularly valuable when the objective function does not have a closed form and is expensive to evaluate—a common scenario in design and engineering where evaluations require physical fabrication or user testing. 

Standard Bayesian optimization proceeds by building a probabilistic surrogate model (typically a Gaussian Process) of the objective function based on past evaluations $\mathcal{D}_t = \{(\design_i, y_i)\}_{i=1}^t$.
From this model, it computes an \emph{acquisition function} $\alpha(\design)$ that scores the utility of evaluating a candidate configuration $\design$.
The optimizer then selects the candidate that maximizes this acquisition function: $\design_{t+1} = \arg\max_{\design} \alpha(\design)$.
A classic choice is \textit{Expected Improvement (EI)} \cite{frazier2018bayesian}, which prioritizes points that are expected to exceed the currently best observed value $y^+$:
\begin{equation}
\ei(\design) = \mathbb{E}_{p(y|\design)}[\max(y - y^+, 0)]
\end{equation}
This approach balances \emph{exploitation} (sampling where the model predicts high performance) and \emph{exploration} (sampling where the model is uncertain).

However, standard formulations of EI implicitly assume that every evaluation incurs the same cost (e.g., the cost $c(\design) \approx \text{const}$).
This cost-blind assumption is problematic in physical prototyping, where the cost difference between tweaking a software parameter and fabricating a new hardware component can be orders of magnitude.
Ignoring these disparities can lead to inefficient resource allocation, where the optimizer wastes budget on expensive configurations that offer only marginal information gains.

To address this, \emph{Cost-Aware Bayesian Optimization} modifies the acquisition strategy to account for heterogeneous evaluation costs.
While various formulations exist---such as additive cost penalties or multi-objective optimization---a robust and effective approach is to optimize for \emph{expected improvement per unit cost} \cite{lee2020cost, snoek2012practical}:

\begin{equation}
\label{eq:cost_acq}
\design^* = \underset{\design}{\arg\max} \quad \frac{\ei(\design)}{\tilde{c}(\design)}
\end{equation}

Here, $\tilde{c}(\design)$ is the estimated cost of evaluating $\design$.
By dividing expected improvement by cost, the acquisition function effectively transforms the search into a rate-of-return problem.
It prioritizes candidates that provide high information gain relative to their expense, encouraging the collection of ``cheap'' information (e.g., via simple parameter tweaks) before committing to ``expensive'' exploration (e.g., full rebuilds), unless the expensive option promises a significantly higher performance breakthrough.

\add{In this work, we adopt this improvement-per-unit-cost formulation. Our specific contribution is not the optimization strategy itself, but the definition of the cost function $\tilde{c}(\design)$ for the domain of interactive device prototyping, which we define in the following sections.}
}

\section{Understanding Cost in Iterative Prototyping of Interactive Devices}
\label{sec:cost}

\begin{figure*}[t]
\centering
\includegraphics[width=0.8\linewidth]{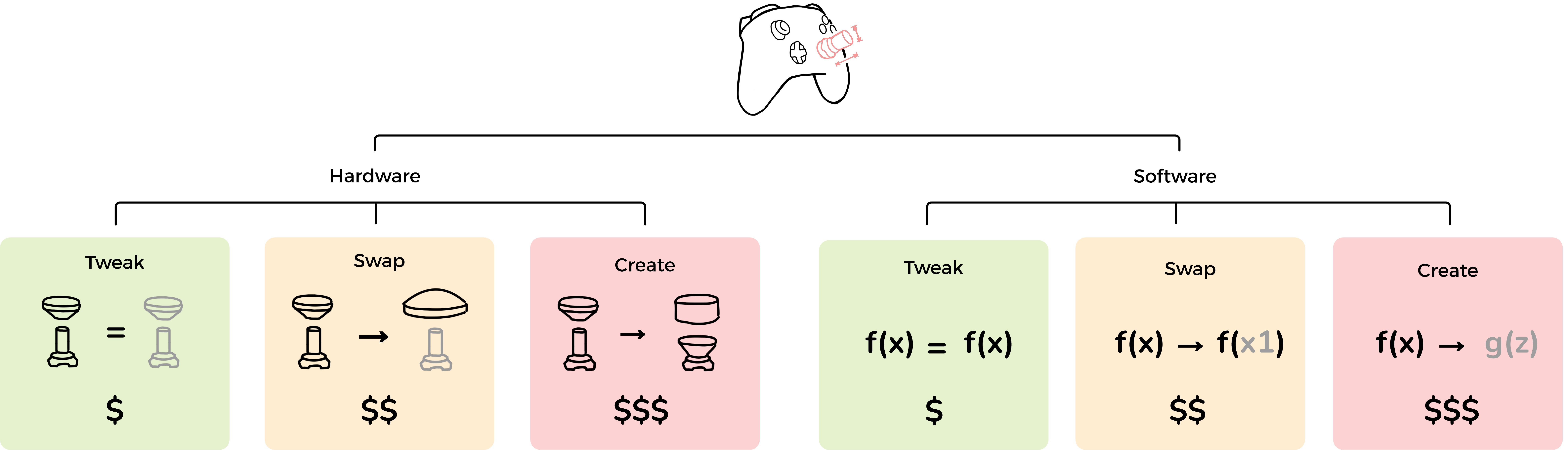}
\vspace{1.0em}
\caption{
\textbf{A classification for costs in iterative prototyping that accounts for already-produced prototypes.} \tweak{} components are unchanged since the last iteration and incur little-to-no cost. \swap{} components have been used in a previous configuration and can be reused at moderate cost. \create{} components have not been built before and must be fabricated or implemented from scratch, incurring the highest cost. Designers fill in these cost types for the cost-aware Bayesian optimization.
}
\Description{Cost structure of hardware and software changes.
A game controller is shown with two branches: hardware modifications on the left and software modifications on the right. Each side distinguishes between tweak, swap, and create operations. Hardware tweaks reuse existing parts at low cost, swaps exchange parts at medium cost, and creates fabricate new parts at high cost. Software tweaks keep the function unchanged at low cost, swaps alter parameters at medium cost, and creates replace functions with new ones at high cost.
}
\label{fig:structure}
\end{figure*}

Prototyping involves diverse costs. For example,  depending on the project, software changes can be inexpensive, while hardware modifications may require materials and hours of fabrication \cite{holmquist2023bits}. These costs often arise at the level of \emph{system components}: discrete elements like mounts, filters, or enclosures, or modules or parameters in software. \add{In this paper, we use “cost” to denote any measurable resource required to evaluate a prototype configuration, such as time, money, labor, or energy.} Costs also depend on what has already been built, what must be modified, and how reusable previous prototype components are.  \add{While any unit of cost may be used, different resources can be converted through simple transfer functions when appropriate (e.g., translating labor or time into monetary terms), though some terms, such as effort, may not convert cleanly.} 
To support more grounded and efficient design iteration, optimization systems must treat cost not as a fixed scalar, but as a structured, evolving constraint tied to past work.

In this section, we examine the costs involved in prototyping interactive systems. We begin with a concrete case that illustrates these costs in practice. We then present a classification of prototyping costs and formalize it as a cost model for Bayesian optimization in the following section. Finally, we extend the model to incorporate a prototype record, capturing opportunities to reuse existing prototypes.

\subsection{Illustrative Case: Prototyping a Joystick}
Consider the iterative design of an Xbox-style joystick for gamers. The designer has identified three main components for improvement: the shaft, the topper, and a software filter. Each can be independently configured: shaft length ranges from, for example, 3–20 mm; toppers vary in width and curvature; and the filter can be switched between smoothing strategies (e.g., fixed gain, adaptive, or none). 

Suppose the designer has already fabricated shafts of 6 mm and 9 mm, toppers with widths of 20 mm and curvatures of 4.0 and 5.0 mm, and implemented a fixed-gain filter. If a new configuration selects a 9 mm shaft, a 20 mm topper with 5.0 mm curvature, and the same filter, all components match earlier builds --- no fabrication is needed.

If instead the designer selects a 6 mm shaft (previously built), a 22 mm topper curvature not yet fabricated, and switches to an adaptive filter not yet implemented, only the shaft can be reused. The topper and filter must be created, designed and implemented from scratch, incurring significantly higher cost. This reflects a common pattern in prototyping: iteration proceeds by selectively building on what already exists.

The cost depends on the designer prioritizes and circumstances; while a print might cost $1\$$ per print in ABS, it could also cause a 3 hour waiting time. A similar trade-off applies to software: while a skilled developer might implement a filter relatively quick (<1h), recompiling firmware for an obsolete controller might take a lot longer (>1 week).

\subsection{Cost Classification}
We introduce a cost classification (\Fig{fig:structure}) that distinguishes three types of costs, spanning both hardware- and software-related aspects. Any unit of cost, such as money, time, or energy, can be used. To formalize this, we assign costs at the level of parameter components: sets of variables that define functional elements of the prototype, such as the shaft or filter. Each component is then categorized based on its relation to previously built configurations:

\begin{itemize}
  \item \textbf{\tweak (\$):} The component is unchanged from the previous iteration and can be reused at minimal cost.  
  \item \textbf{\swap (\$\$):} The component differs from the most recent iteration but appeared in an earlier one, and can be retrieved or reassembled at moderate cost.  
  \item \textbf{\create (\$\$\$):} The component has not been built before and requires full fabrication or implementation at high cost.  
\end{itemize}

This classification reflects how prototyping unfolds: designers avoid full rebuilds unless necessary, reuse known-good configurations when possible, and prefer minimal changes that align with prior prototypes. In \Sec{sec:model} we formalize this classification into a model that can be used by an optimizer.

\subsection{Accounting for Past Prototypes: The Prototype Record}
Design cost is not static: it evolves with iteration. As more configurations are evaluated, the number of reusable components increases, lowering the marginal cost of future changes. To capture this effect, we introduce a prototype record that tracks which parameter components have been built and how recently they were used.

Each configuration is compared against past iterations to determine, per component, whether it is unchanged, reused, or new. This introduces memory into the optimizer, allowing it to exploit the structure of prior prototypes. Rather than treating all changes equally, the system can prioritize those that offer high utility per unit cost.

Our approach captures key properties of real-world prototyping: modular design, selective reuse, and incremental refinement. It remains agnostic to the unit of cost, allowing flexible integration into existing design tools and optimization pipelines.

\subsection{Additional Examples}
We present three additional cases to demonstrate how our cost classification can be applied; for more detail, see \App{app:cases}.

\begin{figure*}[t]
    \centering
    \includegraphics[width=0.8\linewidth]{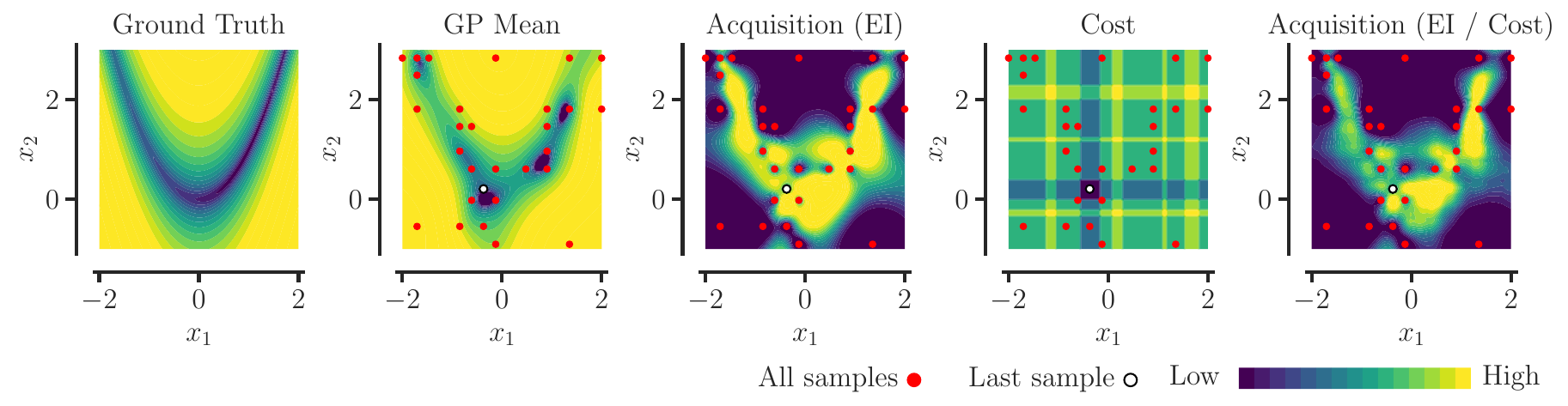}
    \caption{
        \addcr{A snapshot of the iterative optimization process at a single timestep.} Our method continuously cycles through sampling, testing, and updating the model; here, we visualize the components used to select the next sample based on the history of previous samples (red dots). $x_1$ and $x_2$ represent prototype parameters. \textbf{Left:} The Rosenbrock ground truth. \textbf{Center Left:} The Gaussian Process (GP) surrogate model approximating the truth based on current samples. \textbf{Center:} Expected Improvement (EI) derived from the GP. \textbf{Center Right:} The cost landscape for $x_1$ and $x_2$; previously built configurations create low-cost vertical and horizontal bands (biasing the search toward \tweak{} and \swap{} operations). \textbf{Right:} The final cost-aware acquisition function, which scales EI by the cost model. The tolerance $\sigma_g$ has been set high to improve visibility.
        }
    \label{fig:acquisition}
    \Description{Comparison of ground truth, Gaussian Process model, and acquisition functions.
Four heatmaps are shown side by side with axes x1 and x2. The first panel (Ground Truth) displays the true objective surface, with a curved valley. The second panel (GP Mean) shows the Gaussian Process estimate with red dots marking sampled points. The third panel (Acquisition, Expected Improvement) highlights regions with high expected improvement. The fourth panel (Acquisition, Expected Improvement per Cost) adjusts the acquisition by cost, showing altered regions of interest and favoring different sampling areas compared to the standard acquisition. A shared colorbar indicates values from low (dark purple) to high (yellow).
}
\end{figure*}

Dexmo~\cite{gu2016dexmo} is a mechanical exoskeleton for finger tracking and force feedback in VR. Its component structure—finger linkage, force feedback unit (FFU), and controller—maps naturally to our cost classification: software updates such as filtering are low-cost (\tweak), linkage components can often be reused (\swap), and FFU changes require recalibration and refabrication (\create). Relevant costs include fabrication time and material.  

Omni~\cite{langerak2020omni, zarate2020contact} is a volumetric haptic stylus that integrates sensing and actuation. The core electromagnet is expensive to fabricate and simulate (\create), whereas stylus modifications and sensor reconfiguration are lower-cost (\swap), and filter tuning is trivial (\tweak). Beyond time and material, manual labor can be a significant cost.  

Back-Hand-Pose~\cite{wu2020back} uses a wrist-mounted camera and deep learning to estimate hand pose. The key bottleneck is retraining: changes to camera position or data representation often invalidate models (\create), while tweaks to preprocessing (\swap) or learning rate (\tweak) allow partial reuse. Costs specific to this setting include data acquisition and GPU training time.

\section{Method}
\addcr{Our method operationalizes cost-aware optimization for interactive devices.
While the optimization framework (\Eq{eq:cost_acq}) is standard, our primary contribution is the \emph{Cost Model}---the definition of $\tilde{c}(\design)$.} First, prototype parameters are grouped into components (e.g., hardware elements like a shaft or topper, or software functions like a filter). Second, each component is assigned a reuse category that reflects its relation to prior work: unchanged (\tweak), reused from an earlier configuration (\swap), or newly created (\create). Finally, these categories are mapped to numeric costs in any unit (e.g., time, money, energy), which quantify the effort required for each component change.

We integrate this information into the acquisition function used by Bayesian optimization. Instead of selecting the next configuration based solely on expected improvement, we select the one that offers the highest improvement per unit cost (\Eq{eq:cost_acq}). This encourages the optimizer to prioritize low-cost yet informative changes, and to propose costly ones only when they are expected to yield substantial improvements. The result is a plug-and-play replacement for standard Bayesian optimization that increases efficiency in real-world prototyping tasks.

\subsection{Modeling Costs}
\label{sec:model}
In prototyping, cost depends not only on parameter values but also on how those parameters correspond to functional components (\Sec{sec:cost}). Some components must be fabricated or implemented, while others can be reused from earlier iterations. Our cost model makes this distinction explicit by incorporating both component structure and prototype record.

Let $\design$ represent a complete prototype, such as a joystick with specific physical dimensions and software settings. We divide the prototype into components, formally disjoint groups, $G = \{g_1, g_2, \ldots, g_k\}$, where each component $c$ corresponds to a meaningful hardware or software module. Each component includes one or more related parameters, for example, all parameters that handle design, sensors, or a set of software parameters related to input filtering. At each iteration $t$, we track two things: the current prototype $\design_t$ and a prototype record $\mathcal{H}_t = \{G_0, \ldots, G_{t-1}\}$ of all $k$ components that have been fabricated or implemented so far. This record allows us to recognize when a proposed prototype is reusing previous prototypes versus introducing something entirely new. We define the cost $c(\design)$ of evaluating a proposed configuration $\design$ by looking at the status of each component $g \in G$:

\begin{equation}
c_g(\design_g) =
\left\{
\begin{aligned}
& c_{\mathrm{\tweak}}^{g}  && \design_g = \design_{g,t} && \text{(reuse previous prototype)} \\
& c_{\mathrm{\swap}}^{g}   && \design_g \in \mathcal{H}_t, \design_g \neq \design_{g,t} && \text{(reuse from record)} \\
& c_{\mathrm{\create}}^{g} && \design_g \notin \mathcal{H}_t && \text{(create new prototype)}
\end{aligned}
\right.
\end{equation}

The total cost of evaluating $\design$ is the sum of the component costs:
\begin{equation}
c(\design) = \sum_{g \in G} c_g(\design_g).
\end{equation}

Intuitively, this means the sum of a prototype is the sum of all its individual components, regardless of whether they are hardware or software, based on whether they are already configured, need to be swapped, or created. This formulation supports arbitrary components. For example, a joystick may consist of separate hardware components for base, handle, and cap, and software components for filtering or control strategies. Modeling cost at the component level enables reuse-aware optimization that aligns with how cost is distributed in real prototyping workflows.

\subsubsection*{Smooth Relaxation of Cost}
To select the next configuration, Bayesian optimization must optimize the acquisition function. This optimization is typically performed using gradient-based methods, which require all components—including the cost function—to be differentiable. However, our original cost model is discrete: each component is labeled as \tweak, \swap, or \create, with corresponding scalar costs. This makes it incompatible with gradient-based optimization.

To address this, we introduce a smooth relaxation of the discrete cost model. The key idea is to assign each component a \textit{soft} cost by measuring how similar the current configuration is to previously used ones. This allows cost to be treated as a continuous, differentiable function.

We begin by defining a similarity function between parameter values. For each component $g \in G$ (e.g., grip shape, sensor layout), we use a radial basis function (RBF) kernel to quantify similarity between the current design $\design_g$ and a previous one $\design'_g$:

\begin{equation}
K(\design_g, \design'_g) = \exp\left(-\frac{\|\design_g - \design'_g\|^2}{2\sigma_g^2}\right)
\end{equation}

Here, $\sigma_g$ is a tunable parameter controlling sensitivity: smaller $\sigma_g$ values require very close matches for high similarity, while larger values allow broader generalization. We use the RBF kernel because it offers a smooth, tunable measure of similarity that decays with distance, enabling control over how strictly reuse is enforced. Its bandwidth determines whether only near-identical configurations are considered reusable (small $\sigma$) or whether broader generalization is allowed (large $\sigma$). Additionally, the RBF is infinitely differentiable, making it well suited for gradient-based optimization.

We compare the proposed configuration $\design_g$ against two references:
1) the current configuration $\design_{g,t}$, which defines whether a component is unchanged (\tweak) and 2) the prototype record $\mathcal{H}_g = \{\design_{g,0}, \ldots, \design_{g,t-1}\}$, which stores all previous configurations for component $g$ (enabling \swap).

We compute soft similarity weights using RBF kernels:
\begin{align}
w_{\text{\tweak}} &= K(\design_g, \design_{g,t}) \\
w_{\text{\swap}}  &= \sum_{\design'_g \in \mathcal{H}_g} K(\design_g, \design'_g) 
\end{align}
The weight for \create, $w_{\text{\create}}$, is treated as a tunable parameter as it cannot be computed from similarity. This ensures the model always retains a nonzero baseline probability of proposing novel configurations, regardless of kernel similarity. Setting $w_{\text{\create}}$ to a higher value encourages exploration by favoring unseen prototypes with higher fabrication cost. Lower values make the optimizer more conservative, focusing on reuse and small adjustments. In practice, this parameter controls how aggressively the optimizer is allowed to deviate from existing configurations.

We compute a soft cost for each component by defining similarity-based weights using RBF kernels. These weights are normalized to produce a probability distribution over reuse types, which we use to interpolate the component cost:

\begin{equation}
\tilde{c}_g(\design_g) =
\frac{w_{\text{\tweak}} \cdot c_{\text{\tweak}}^g +
w_{\text{\swap}} \cdot c_{\text{\swap}}^g +
w_{\text{\create}} \cdot c_{\text{\create}}^g}
{w_{\text{\tweak}} + w_{\text{\swap}} + w_{\text{\create}}}
\end{equation}

The total cost is the sum over all components:
\begin{equation}
\tilde{c}(\design) = \sum_{g \in G} \tilde{c}_g(\design_g)
\end{equation}

Our formulation provides a smooth, differentiable relaxation of the discrete cost model while retaining its core intuition: configurations closer to existing or current designs incur lower cost. Tunable weights and the bandwidth parameter $\sigma_g$ allow control over reuse sensitivity. Narrow kernels (small $\sigma_g$) enforce near-exact reuse, while wider kernels (large $\sigma_g$) permit more flexible similarity.

This relaxation makes cost-aware acquisition functions compatible with gradient-based optimization. It enables efficient integration into modern Bayesian optimization pipelines while still respecting the structure and intent of the original reuse-based model.

\add{In summary, the proposed cost classification is in discrete space (either \tweak, \swap or \create). To be able to optimize the acquisition function we need to ensure smooth gradients. To that end, we transform this discrete space into a continuous space. We do so by approximating every discrete cost by RBF kernels based on the similarity between the proposed prototype and the prototype record. Thus, this similarity weighting is a relaxed version of the proposed cost structure. }

\subsubsection*{Updating the Prototype Record}
After evaluating a new configuration $\design_t$, we update the prototype record $\mathcal{H}t$ to reflect any components that were newly created. Formally, for each component $g \in G$, if $\design_{g,t} \notin \mathcal{H}_{g}$, we append it to the record:
\begin{equation}
\mathcal{H}_{g} \leftarrow \mathcal{H}_{g} \cup {\design_{g,t}} \quad \text{if } \design_{g,t} \notin \mathcal{H}_g.
\end{equation}

\begin{figure*}[t]
    \centering
    \includegraphics[width=0.8\linewidth]{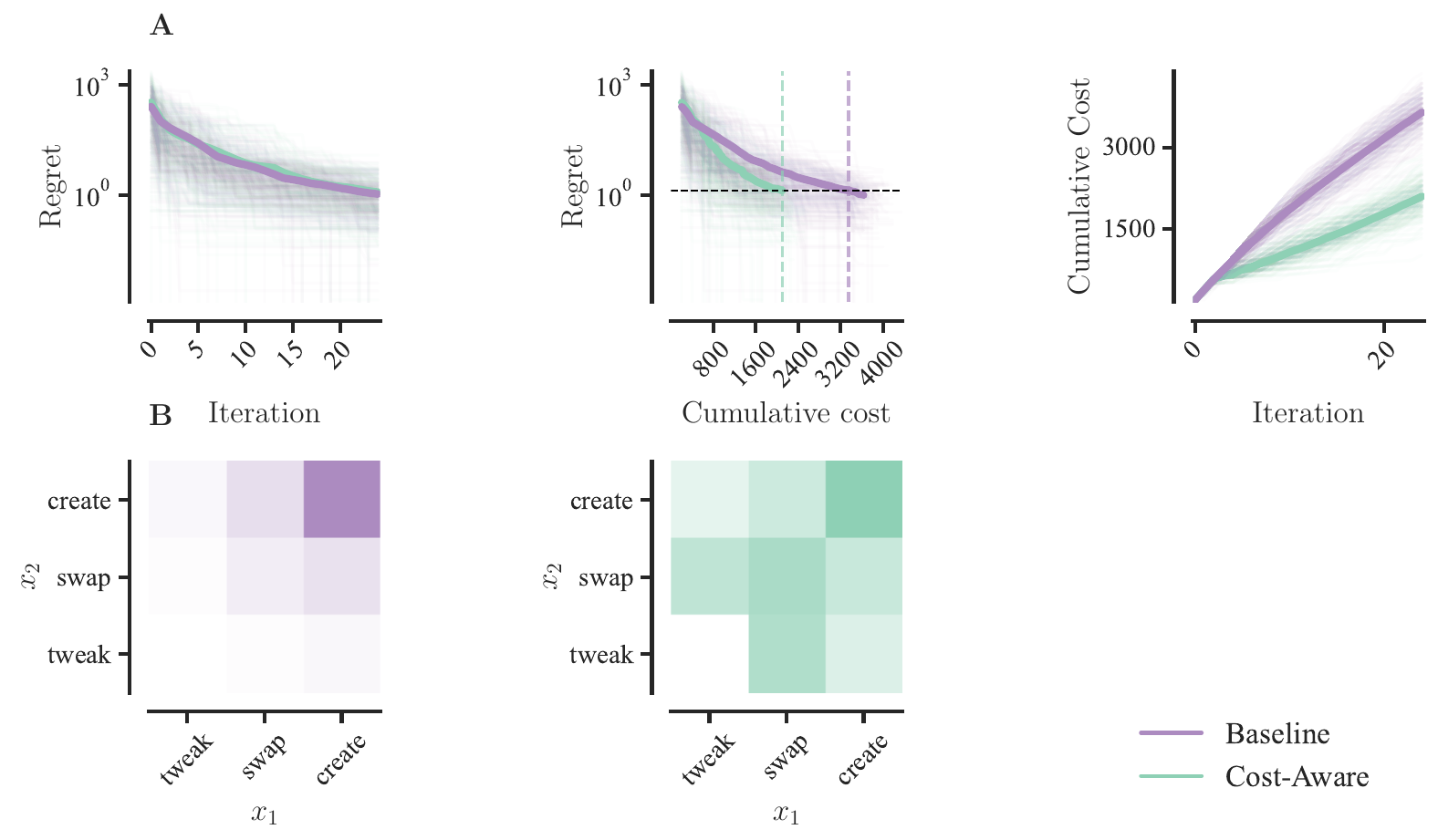}
\caption{\textbf{Study 1.} Cost-aware vs.\ baseline BO with fixed iterations. \textbf{A:} Performance metrics. \textit{Left:} Baseline converges to slightly lower regret. \textit{Center:} Cost-aware achieves similar utility at $\sim$2/3 the cost. Vertical lines mark the cost-aware minimum. \add{\textit{Right:} Baseline accumulates cost significantly faster.} (Means exclude runs with $<95\%$ trials). \textbf{B:} Sample counts per hardware ($x_1$) and software ($x_2$) operation. Baseline over-samples expensive \textit{\create} edits, while cost-aware distributes samples evenly across cost structures.}
    \label{fig:eval1}
    \Description{Performance comparison of baseline and cost-aware Bayesian optimization.
Panel A shows two line graphs of regret. The left graph plots regret against iteration, with both baseline (purple) and cost-aware (green) optimizers steadily decreasing over 25 iterations. The right graph plots regret against cumulative cost. The cost-aware method reaches low regret earlier and with lower cost, indicated by dashed vertical lines. Panel B shows heatmaps of sampling frequency over tweak, swap, and create categories. The baseline method (right heatmap) samples more costly create options, while the cost-aware method (left heatmap) favors cheaper tweaks and swaps.
}
\end{figure*}

This update ensures that future cost calculations recognize the availability of reusable components. The prototype record thus evolves with the optimization process, enabling the system to avoid redundant fabrication and to prefer configurations that leverage prior effort. 

\begin{algorithm}[b]
\caption{Cost-Aware Optimization with Smooth Relaxation}
\label{alg:opt}
\begin{algorithmic}[1]
\Require Parameter space $\mathcal{X}$ with groups $G$, GP prior $\mathcal{GP}_0$, initial prototype record $\mathcal{H}_0$, cost levels $\{c_{\tweak}^g, c_{\swap}^g, c_{\create}^g\}$ for each group $g$
\State $t \gets 0$, $\mathcal{D}_0 \gets \varnothing$, $\mathcal{H}_t \gets \mathcal{H}_0$, $\mathcal{GP}_t \gets \mathcal{GP}_0$
\While{not converged}
    \State \textbf{Define acquisition:}
    \Statex \quad $a_{\mathrm{cost}}(\design) = \ei_{\mathcal{GP}_t}(\design) \,/\, \tilde{c}(\design, \mathcal{H}_t)$
    \State \textbf{Optimize acquisition:}
    \Statex \quad $\design^\ast \gets \arg\max_{\design \in \mathcal{X}} a_{\mathrm{cost}}(\design)$
    \State \textbf{Execute and evaluate:}
    \Statex \quad Realize configuration $\tilde{\design}$ (e.g., snapping to tolerance)
    \Statex \quad Compute discrete cost $c^\ast \gets c(\tilde{\design}, \mathcal{H}_t)$
    \Statex \quad Observe utility $y^\ast = f(\tilde{\design})$
    \State $\mathcal{D}_{t+1} \gets \mathcal{D}_t \cup \{(\tilde{\design}, y^\ast)\}$
    \State $\mathcal{H}_{t+1} \gets \mathcal{H}_t \cup \textsc{NewComponents}(\tilde{\design}, \mathcal{H}_t)$
    \State $\mathcal{GP}_{t+1} \gets \textsc{UpdateGP}(\mathcal{GP}_t, \mathcal{D}_{t+1})$
    \State $t \gets t+1$
\EndWhile
\end{algorithmic}
\end{algorithm}

\subsection{Algorithm Walkthrough}
This algorithm follows the standard Bayesian optimization loop but modifies the acquisition function to account for cost. At each iteration (\Alg{alg:opt}), the system selects a configuration that offers the best expected improvement relative to its estimated cost, using the smooth relaxation described above. This encourages reuse and small adjustments unless a new configuration is predicted to yield substantial improvement. Once a configuration is selected, it is snapped to the nearest buildable prototype (e.g., to match fabrication constraints), evaluated, and added to the dataset used to update the GP. The prototype record is also updated to reflect any newly created components, which influences future cost estimates. Over time, this process helps allocate prototyping effort more efficiently by prioritizing low-cost but informative changes.

\section{Technical Evaluation}
\label{sec:technical}
To evaluate whether our cost-aware optimization method improves efficiency in settings that reflect real-world prototyping, we conduct five simulation studies. These experiments isolate and study key challenges in iterative design: cumulating costs, budget constraints, asymmetric cost structures, parameter complexity, and dynamic resource availability. Together, these tests assess whether explicitly modeling cost improves optimization under realistic conditions.

\subsection{Baseline and Experimental Setup}
We compare our approach to a standard Bayesian optimization method common in HCI applications~\cite{liao2023human}. The baseline selects candidates via expected improvement, without using cost. Our method modifies the acquisition function to instead prioritize expected improvement per unit cost.

To enable systematic benchmarking, we use the 2D Rosenbrock function as a ground-truth utility surface (see \Fig{fig:acquisition}):
\begin{equation}
    f(x_h, x_s) = (1 - x_1)^2 + 100(x_1 - x_2^2)^2
\end{equation}
where $x_1$ and $x_2$ different prototype parameters, such as hardware and software. The Rosenbrock function is commonly used in optimization due to its narrow, curved valley~\cite{shampine1982implementation}. It also serves as a proxy for HCI prototyping as the interdependence of \(x_1\) and \(x_2\) mirrors hardware–software coupling. \add{We evaluate three more (Ackley, Goldstein-Price and Levy) ground truth functions in \App{app:extragt}.}

To simulate human-in-the-loop evaluation noise, we add both additive and multiplicative Gaussian noise to the function output:
\begin{equation}
    y = f(x)*\epsilon_m + \epsilon_a,\quad
    \epsilon_a \sim \mathcal{N}(0, 0.1),\quad
    \epsilon_m \sim \mathcal{N}(1, 0.1)
\end{equation}
Additive noise models fixed evaluation variability such as hand jitter, while multiplicative noise reflects how measurement uncertainty scales with output magnitude.

Cost is structured across components and actions. Each parameter belongs to a hardware or software component. Changes incur cost based on action type: \tweak = 1, \swap = 10, \create = 100, for both components.

Optimization runs begin with three random configurations drawn via Sobol sampling. Both methods use the same GP model, acquisition optimizer (L-BFGS-B), and stopping criteria. All parameters are normalized to $[0,1]$ for modeling, and de-normalized for execution.

Performance is measured using regret --- the difference between the global optimum and the best observed value --- capturing both convergence speed and solution quality. We report regret over iterations to assess sample efficiency and regret over cumulative cost to assess practical efficiency.

\paragraph{Implementation} Our implementation is written in Python 3.12 using scikit-learn~\cite{pedregosa2011scikit} for Gaussian Process modeling and SciPy~\cite{virtanen2020scipy} for numerical optimization. All experiments were conducted on a Apple Macbook M3, and run in real-time. To ensure robustness all experiments include multiple trials initialized with random seed. Source code is available at \url{https://github.com/aalto-ui/CABOP}.

\subsection{Study 1: Performance Under Unlimited Budget}
Designers often iterate until a design is good enough \cite{wynn2017perspectives}. In this setting, cost still matters: it determines the total process cost of reaching the final design. This study tests whether our cost-aware method achieves similar outcomes with less cumulative cost, even when explicit budget is imposed.

\paragraph{Setup}  
We optimized the 2D Rosenbrock function over 25 iterations with 250 independent trials per method. Both methods began with three random Sobol samples. At each step, we recorded the regret and the action cost type (\tweak, \swap, \create). While iteration count was fixed, cumulative cost was not, allowing us to compare sample efficiency (steps) and cost efficiency (resources used). Statistical analysis used Mann–Whitney U tests due to non-normality of distributions.

\paragraph{Results}  
Results are shown in \Fig{fig:eval1}. Over 25 iterations, both methods reduced regret, with the baseline achieving slightly lower final regret ($M = 1.01$, $SD = 1.18$) than the cost-aware method ($M = 1.14$, $SD = 1.17$). However, this difference was not statistically significant ($p = 0.12$). Our method also reached its best regret at lower cost ($M = 1529$, $SD = 542$) than the baseline ($M = 2785$, $SD = 890$), $p < .01$. \add{The final cost in our approach was also markedly lower than in the baseline (cost: $M = 2101.39$, $SD = 277.91$; baseline: $M = 3642.76$, $SD = 290.41$), a difference that was significant ($U = 29.50$, $p < .01$).} Sampling behavior differed significantly: the baseline heavily overused expensive \create actions, while our method favored cheaper edits. A chi-squared test confirmed this pattern ($\chi^2(5, N = 26000) = 2956.87$, $p < .01$), with large standardized residuals indicating strong over-selection of costly actions by the baseline, while ours uses lower-cost \tweak and \create, without less of result. This results indicate that our method utilizes cheaper options, to lower the cumulative cost, without compromising on quality.  

\subsection{Study 2: Performance Under Budget Constraints}
\begin{figure}[t]
    \centering
    \includegraphics[width=\linewidth]{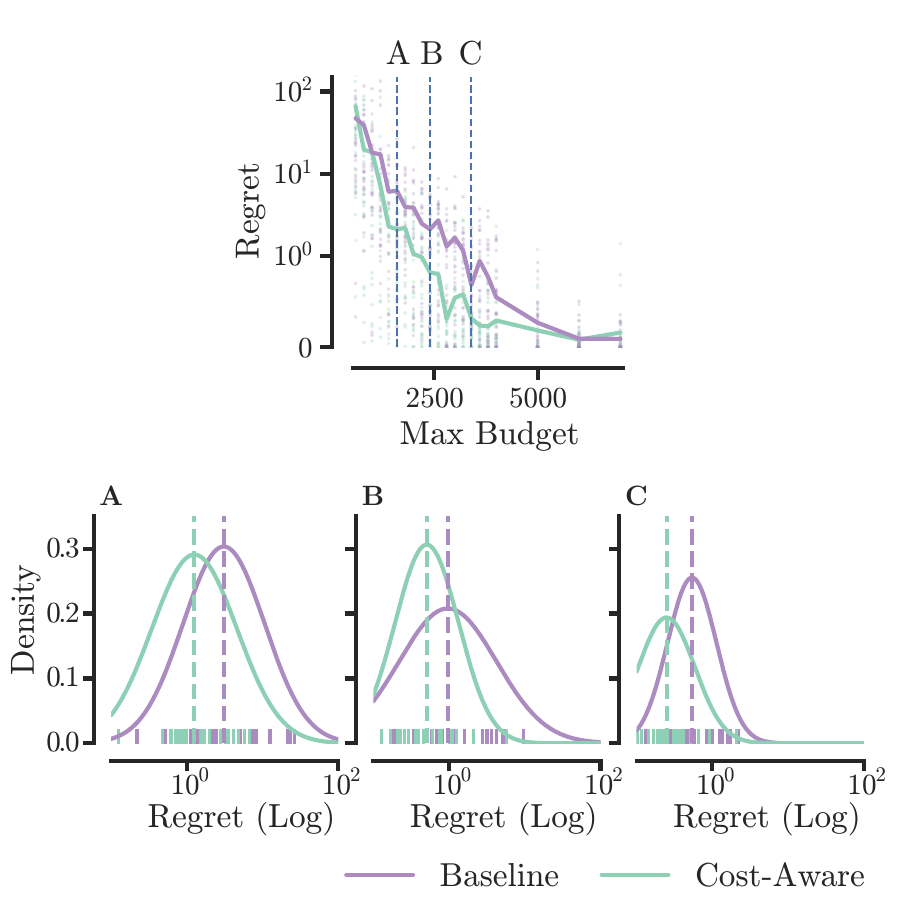}
    \caption{\textbf{Study 2.} \textbf{Tp[]:} Final regret and regret distributions across fixed cost budgets ranging from 600 to 7000 units. \textbf{Others}: Cross-section at specific budgets, corresponding to the vertical lines in the left plot. The cost-aware approach significantly outperforms the baseline in the middle segment, where the total budget is not negligible but also not inconsequentially big.}
    \label{fig:eval2}
    \Description{Regret distributions under different budget levels.
 The left panel plots regret versus maximum budget. Both baseline (purple) and cost-aware (green) optimizers reduce regret as budget increases, with cost-aware achieving lower regret earlier. The three right panels show density plots of regret at budget levels 1600, 2400, and 3400. At each level, the cost-aware method’s distribution (green) is shifted toward lower regret compared to the baseline, indicated by dashed vertical lines and tighter distributions.}
\end{figure}
Most real-world prototyping happens under resource limits \cite{tiong2019economies}. Designers rarely optimize freely; instead, they stop when funds, time, or materials run out. This study evaluates whether our method achieves better results when the total available budget is capped.

\paragraph{Setup}  
We simulated optimization under fixed cost budgets ranging from 600 to 7000 units. Each method used its entire budget and terminated once exhausted. For each budget level, we recorded the final regret. At each level, 25 optimization runs were performed per method. As in Study 1, we used Mann–Whitney U tests to assess statistical significance due to non-normal outcome distributions.

\paragraph{Results}
Results are shown in \Fig{fig:eval2}. Across most of the budgets, the cost-aware method consistently achieved lower final regret, with the most significant gains observed at mid-range budgets. At a budget of 1600, regret was significantly lower for our method ($M = 2.11$, $SD = 1.83$) than for the baseline ($M = 6.21$, $SD = 7.48$), $p = .027$. At 2400, this gap widened further ($M = 0.80$ vs.\ $2.70$, $p < .01$). Even at 3400, when both methods approached convergence, cost-aware optimization still performed better ($p = .02$). This shows a balance, if there is no budget virtually all needs to be spend on prototype creation, as in the early stages there is no prototype record to rely on, while if the budget is sufficiently big both methods are able to find an optimal design. Our method offers the greatest advantage in the intermediate regime, where efficient reuse and selective creation lead to better outcomes under limited resources.

\subsection{Study 3: Sensitivity to Cost Asymmetry}
\begin{figure}
    \centering
    \includegraphics[width=\linewidth]{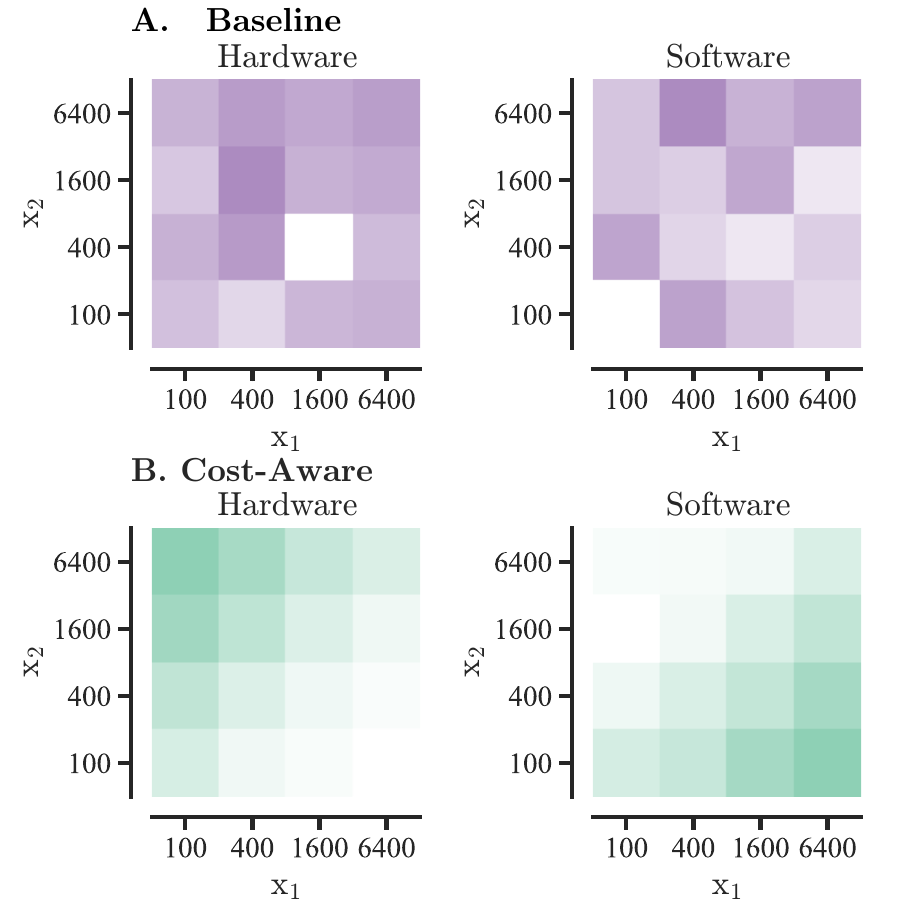}
    \caption{\textbf{Study 3.} Sample counts by method, component type, and cost asymmetry for \create, while the other costs stay constant. Each heatmap shows the number of hardware or software samples as a function of hardware and software \create cost. Our method adapts its sampling behavior to cost asymmetries, reducing selection of expensive components, while the baseline samples each component uniformly. Heatmaps are individually normalized for readability.}
    \label{fig:eval3}
    \Description{Sampling patterns for hardware and software parameters.
Four heatmaps compare baseline (purple, left two panels) and cost-aware (green, right two panels) optimizers. Each heatmap shows sampling frequency across parameter values x1 and x2. In hardware (first and third panels), the baseline samples broadly, while the cost-aware optimizer favors lower-cost regions. In software (second and fourth panels), the baseline samples evenly across settings, while the cost-aware optimizer emphasizes cheaper parameter combinations.
}
\end{figure}

Prototyping often involves components with very different costs \cite{christie2012prototyping}. In particular, hardware changes typically incur greater fabrication time and material use compared to software edits. This study evaluates whether our method adapts to such cost asymmetries, whereas standard approaches treat all evaluations uniformly.

\paragraph{Setup}  
We independently varied hardware and software \create costs in \{100, 400, 1600, 6400\}, yielding 16 cost conditions. For each condition, we ran 50 optimization trials on the 2D Rosenbrock function, each for 25 iterations. We recorded the number of samples taken by component and action type (\tweak, \swap, \create). A generalized linear mixed-effects model assessed whether the methods adjusted sampling behavior based on cost asymmetries.

\paragraph{Results}  
Our method shifted its sampling strategy (see \Fig{fig:eval3}) in response to changing costs: as hardware cost increased, it performed significantly fewer hardware \create actions and more low-cost software edits ($p < .01$ for interaction effects). The baseline showed no such adaptation --- hardware and software were sampled at nearly constant rates, regardless of cost. Post-hoc contrasts confirmed that the baseline over-sampled hardware \create (estimate = 0.62 on the log scale, $p < .01$) and under-sampled software edits. In contrast, our method redistributed effort toward cheaper software parameters when hardware costs were high. This indicates that only the cost-aware method dynamically shifts effort across modalities in response to asymmetry, aligning better with real-world constraints.

\subsection{Study 4: Scaling with Prototype Complexity}

\begin{figure*}[t]
\centering
\includegraphics[width=0.8\linewidth]{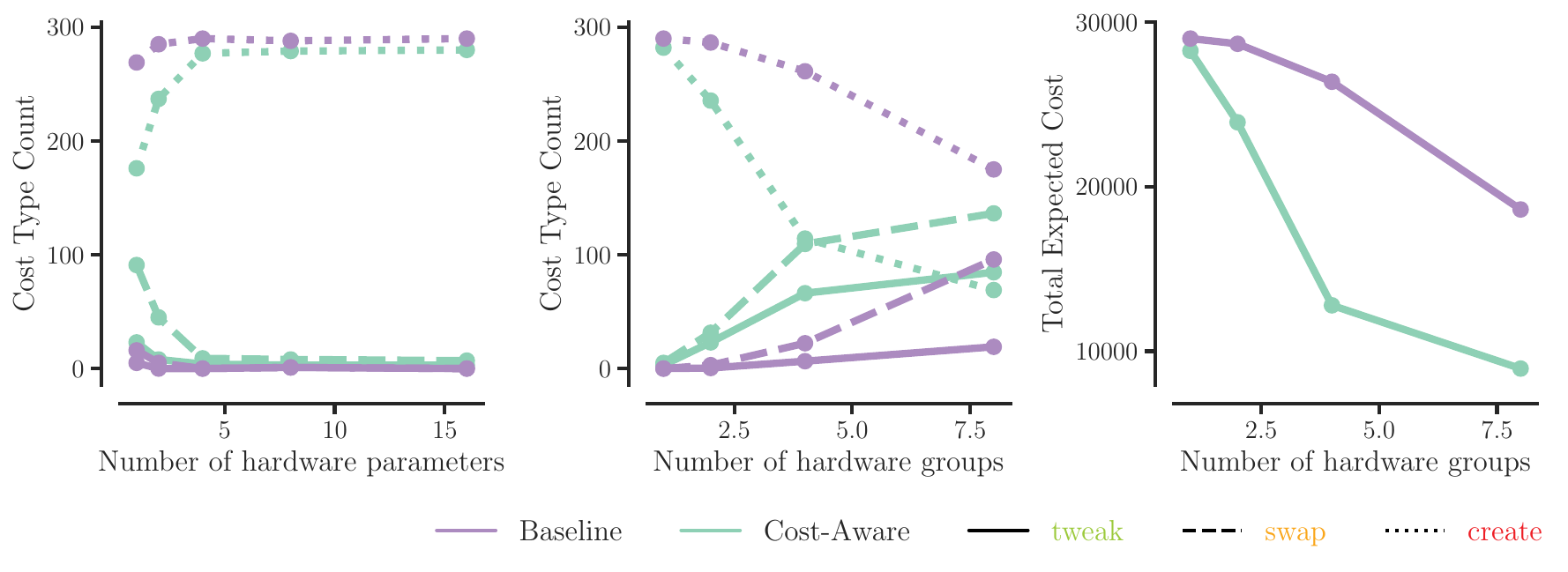}
\caption{\textbf{Study 4.} Count of iterations where hardware parameters are \tweak, \swap, or \create, as a function of dimensionality. The number of parameters negatively impacts both methods, however, when the number of components increases the cost-aware method gains significantly more benefit compared to the baseline. \add{This study was using the $n$-dimensional Rosenbrock function.}}

\Description{Effect of cost-aware optimization on hardware cost distribution and total cost.
Three line graphs compare cost-aware (green) and baseline (purple) optimizers. The first panel plots cost type distribution against the number of hardware parameters, showing the baseline dominated by expensive create operations, while cost-aware favors tweaks and swaps. The second panel plots cost type distribution against the number of hardware components, again showing the baseline skewed toward creates, while cost-aware balances across cost types. The third panel plots total expected cost, where cost-aware consistently achieves lower costs as the number of components increases.
}
\label{fig:eval4}
\end{figure*}

As prototypes become more complex, designers face a growing number of parameters and components \cite{staib2013components}. Some systems have dense coupling, where modifying any parameter requires full rebuilds, while others are decomposed into distinct components, allowing localized edits. This study evaluates whether our method scales effectively with increasing complexity and exploits components to reduce cost.

\paragraph{Setup}  
We evaluate two forms of complexity using the multi-dimensional Rosenbrock function. First, we varied the number of parameters (\{1, 2, 4, 8, 16\}), which were in a singular component and thus any change more likely triggered a full \create action. Second, we fixed the total number of hardware parameters to 8 and varied the number of independently fabricated components (\{1, 2, 4, 8\}), with changes only affecting the corresponding component. For both, we ran 50 trials of 25 iterations and analyzed sampling behavior using generalized linear mixed-effects models.

\paragraph{Results} 
\add{The results are shown in \Fig{fig:eval4}}, when parameter count increased within a single component, both methods relied more heavily on costly \create actions. However, our method limited this increase more effectively at low to moderate dimensionalities. For example, at 1–2 parameters, \create actions were significantly lower in our method than the baseline ($p < .05$), but this difference diminished at higher dimensions ($p > .5$). This suggests that both methods suffer from increased parameter coupling.
In contrast, when modularity increased, our method consistently reduced \create sampling and increasingly relied on low-cost edits. The benefit emerged from \texttt{NComponent} = 2 onward ($p < .05$) and grew with further modularity, with a final difference of over 10 \create actions on average at \texttt{NComponent} = 8. The baseline showed a lesser effect. These results show that our method exploits structural opportunities for reuse, while the baseline only minor adjusted its sampling to system structure.

\subsection{Study 5: Reweighting the Costs}

Real-world prototyping is dynamic: fabrication resources, material access, or team expertise may change mid-process \cite{christie2012prototyping}. These fluctuations alter the effective cost of design actions. This study tests whether our method adapts its sampling behavior when costs change during a single optimization cycle.

\paragraph{Setup}  
We evaluate two conditions over 24 iterations with three random initializations. In the \textit{constant} condition, cost weights remain fixed. In the \textit{dynamic} condition, hardware \create cost increases tenfold at iteration 10 (e.g., tool unavailability), and then drops to one-tenth at iteration 17 (e.g., faster fabrication found). We compare \create action frequency across three cost phases: iterations 4–10, 11–17, and 18–24. Each condition was repeated for 50 optimization trials. A generalized linear mixed-effects model assessed the interaction between condition, phase, and action type, focusing on \create sampling.

\paragraph{Results}  
Our method adapted strongly to cost changes. In the dynamic condition, \create sampling dropped significantly after the cost increase ($p < .01$) and rose sharply when cost decreased again ($p < .01$). In contrast, the constant condition showed no meaningful shifts. Between conditions, the dynamic run sampled fewer \create actions during high-cost phases ($p = .01$) and more when costs dropped ($p < .01$). These changes occurred despite identical utility surfaces, demonstrating that our method tracks cost structure, not just outcome. These results show that our method dynamically reallocates sampling effort in response to shifting constraints. 

\add{
\subsection{Study 6: Effect of Inaccurate Cost Estimate}
\begin{figure*}
    \centering
    \includegraphics[width=0.8\linewidth]{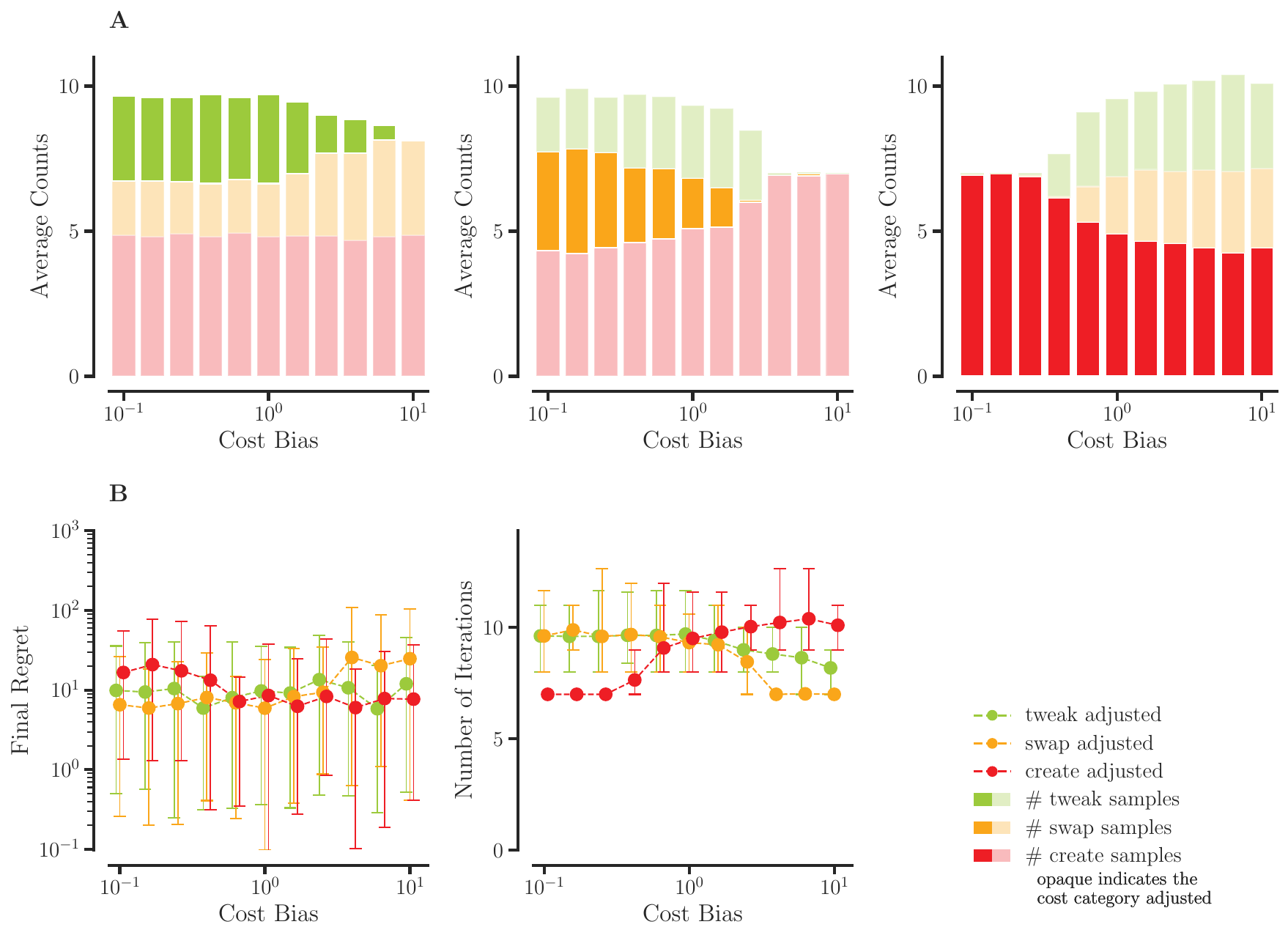}
    \caption{\add{\textbf{Study 6:} Effect of inaccurate cost estimates on optimization behavior. The believed cost is scaled relative to the true cost by a factor $\alpha$, ranging from $10^{-1}$ to $10^{1}$. \textbf{A:} sampling behavior across \tweak, \swap, and \create actions as function of which cost category has a bias (opaque). \textbf{B}: The plots show how mis-specified costs influence final regret (left) and iteration count (center), illustrating overall the robustness of the method and the complex interplay costs have. Error bars indicate a 95\% confidence interval. 
    Opaque bars mark the cost category that was adjusted; light bars show the unchanged categories. Each line in B corresponds to the same cost adjustment in A.}}
    \Description{Effect of cost bias on regret, iterations, and sampling behavior across tweak, swap, and create actions.
The top row shows two line plots across bias levels. Final regret varies moderately with bias, and the number of iterations remains nearly constant, with small differences depending on which action type is cost-adjusted.
The bottom row contains three bar charts showing average sampling counts when each action type is cost-adjusted. Adjusting tweak reduces tweak samples at high bias; adjusting swap reduces swap samples; and adjusting create reduces create samples at low bias but increases them when bias favors creating. These plots illustrate how cost bias shifts the relative frequency of sampled actions.
}
    \label{fig:eval8}
\end{figure*}
In our final experiment, we challenge the method's robustness to inaccurate cost specifications---a common real-world scenario where designer estimates diverge from actual resource consumption. This presents a difficult optimization problem: the method must navigate the utility landscape using a cost model that acts as a ``wrong map,'' decoupling the optimizer's internal logic from the strict external budget constraints that terminate the episode.

\paragraph{Setup}
We introduce a discrepancy between the \textit{believed cost} (used by cost-aware Bayesian optimization for planning) and the \textit{true cost} (deducted from the finite budget). We define the believed cost as the true cost scaled by a factor $\alpha$, ranging from $10^{-1}$ (severe underestimation) to $10^{1}$ (severe overestimation) on a log scale. The optimizer assumes the believed cost is accurate and does not update it during the run. We evaluate the impact of this mismatch on final regret, iteration count, and the change in sampling behaviors (\tweak, \swap, and \create). We ran 50 randomized trials.

\paragraph{Results}
\Fig{fig:eval8} shows that the overall pattern is consistent: increasing the believed cost of a prototype cost reduces its use, and decreasing it increases sampling. Depending on the category this influences the number of iterations and final regret.

The effects differ by category. When \tweak is assumed expensive, the optimizer shifts toward \swap; when assumed cheaper, the impact is small because it has no lower–cost rival. Misestimating \swap has stronger impacts. Underestimation increases \swap use, replacing \create and raising the iteration count; overestimation pushes the optimizer toward \create, reducing iterations and increasing regret. For \create, the effect is direct: underestimation causes overuse and faster budget drain, while overestimation shifts sampling to \tweak and \swap, increasing iterations.

Overall, the method adapts sampling in a predictable way while preserving performance. The impact depends on the operator and the direction of the error, suggesting that maintaining proportional cost relationships is more important than exact values. 
}
\subsection{Summary}

\begin{table}[t]
    \centering
    \small
    \begin{tabularx}{\columnwidth}{@{} >{\raggedright\arraybackslash}X c c @{}}
        \toprule
        \textbf{Study} & \textbf{Standard} & \textbf{Cost-Aware} \\
        \midrule
        \textbf{S1:} Cost efficient & \xmark & \cmark \\
        \textbf{S2:} Effective under cost budget constraints & \xmark & \cmark \\
        \textbf{S3:} Adapts to cost asymmetries & \xmark & \cmark \\
        \textbf{S4:} Exploits modular structure & $\sim$ & \cmark \\
        \textbf{S5:} Responds to changing cost mid-cycle & NA & \cmark \\
        \textbf{S6:} Sensitivity to cost uncertainty & NA & \cmark \\
        \bottomrule
    \end{tabularx}
    \caption{Comparison of the baseline and our cost-aware approach across six studies. The cost-aware approach consistently outperforms the baseline. $\sim$: While the standard approach had an effect with the modular structure, this is inherent to increased components and not due to the method.}
    \label{tab:techeval}
    \Description{Comparison of capabilities supported by standard versus cost-aware optimization.
The table lists five capabilities across studies S1 to S5. Cost-aware optimization supports all five capabilities. Standard optimization does not support cost efficiency (S1), effectiveness under cost budget constraints (S2), or adaptation to cost asymmetries (S3). For exploiting modular structure (S4), standard optimization provides partial support. For responding to changing cost mid-cycle (S5), no assessment is available.}
\end{table}

Results are summarized in \Tab{tab:techeval}.
Overall, we learn that modeling cost structure enables smarter sampling and better resource use.
In Study~1, the cost-aware method matched the baseline in performance but used significantly less cumulative cost. Study~2 showed stronger gains under fixed budgets, where cost-aware optimization achieved lower regret by allocating effort more efficiently. Study~3 confirmed that our method adapts to cost asymmetries, shifting effort toward cheaper edits; mirroring how designers prioritize low-effort changes. In Study~4, only the cost-aware method leveraged modularity to reduce expensive rebuilds, showing structural sensitivity absent in the baseline. Study~5 demonstrated dynamic adaptation: when costs shifted mid-cycle, our method adjusted sampling accordingly, maintaining efficiency throughout. \add{Finally, Study~6 showed that the method remains robust even under  mis-specified costs, adjusting sampling in a principled direction and preserving overall performance despite inaccurate estimates.}

\paragraph{Evaluation Robustness}
While we use the Rosenbrock function, we do not claim it captures the full structure of HCI design spaces. However, to test robustness, we ran all simulations with random seeds and noise. Our results suggest that the performance gains are not fragile to noise or initial conditions, and that the observed behavior is not an artifact of a specific random run. \add{Additional results using alternative ground-truths can be found in \App{app:extragt}.}

\section{User Study}
\label{sec:user}
\begin{figure*}
    \centering
    \includegraphics[width=0.75\linewidth]{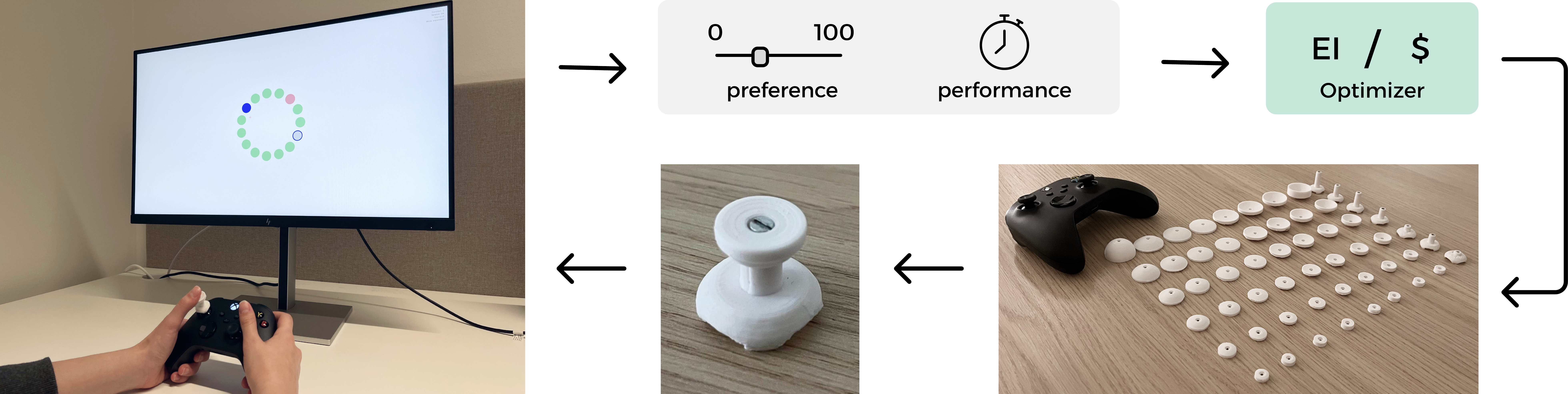}
    \vspace{1.0em}
    \caption{
        User study overview.  
        Participants perform a Fitts's law task, where they use a joystick to move a cursor toward a highlighted target (blue filled circle) and confirm the selection using either the A-button or one of the triggers. Upon successful selection, the target reappears at the opposite side of the ring and the task repeats.
        After each trial, participants rate the tested configuration on a scale from 0 to 100.  
        The optimizer integrates both subjective preference and objective performance to propose a new configuration.  
        The suggested configuration is swapped into the device, and the process repeats until the number of iterations are reached.
    }
    \Description{User study setup with joystick prototypes and cost-aware optimization loop.
 The left photo shows a participant holding a joystick while performing a pointing task on a monitor with circular targets. On the right, a flow diagram illustrates the optimization loop: user preference ratings and performance times feed into an optimizer based on expected improvement per cost. The optimizer selects new joystick prototypes, shown in the lower photo as a set of interchangeable 3D-printed caps arranged on a table alongside the controller.}
    \label{fig:task}
\end{figure*}

We conducted a within-subjects user study to evaluate whether cost-aware human-in-the-loop optimization improves efficiency in customizing a hardware–software joystick configuration, compared to a standard Bayesian optimization baseline. This testbed was chosen because it is realstic, and it combines actual hardware edits that are slow and costly with software changes that are quick and inexpensive, creating realistic asymmetries in evaluation effort. 

The optimizer combined task performance with participants’ comfort ratings. Before optimization began, each participant specified a weighting between performance and preference, which was then used throughout the session to compute the utility. This measure allowed us to evaluate whether our approach could reduce prototyping effort without degrading the quality of outcomes.

\paragraph{Participants}
We recruited 12 participants (5 female, 7 male; age: $M = 23.3$, $SD = 3.3$) through local advertisements. All participants had normal or corrected-to-normal vision and reported daily computer use. Participants were screened for prior gaming experience to ensure sufficient hand–eye coordination. Handedness was not a factor. All participants completed the study. No data were excluded. Each received a €20 voucher for a local restaurant as compensation.

\paragraph{Task \& Materials}

The joystick design was based on Adaptive Thumbstick Toppers from Microsoft Xbox \cite{xboxAdaptiveThumbstick}. These are 3D-printable modifications to standard controller thumbsticks. Participants customized both hardware (shaft length, topper convexity, topper width) and software parameters (sensitivity, and reactivity), summarized in \Tab{tab:userstudyparas}. To make the task feasible within the experiment time a large variety of the hardware parameters was fabricated \emph{a priori} (over 600 combinations of shaft and topper). 

The cost structure for both sensitivity and reactivity was set to {\tweak: 1, \swap: 10}. Since all software configurations were pre-implemented for runtime selection, \create costs were not applicable. Both software parameters were treated as independent components. For hardware, the shaft was assigned costs of {\tweak: 1, \swap: 10, \create: 100}, while the topper was set to {\tweak: 1, \swap: 10, \create: 1000}. The higher cost of creating a new topper reflects its increased physical complexity and longer 3D printing time. These values were chosen to approximate relative effort rather than exact timing --- for example, software changes typically involve navigating menus, while hardware fabrication, although low in manual labor, incurs long delays due to print time. The aim was to model the asymmetry and practical friction of real-world design iteration without enforcing a strict mapping to absolute duration.

The task was reciprocal circular Fitts's law task, consisting of 15 targets at two index of difficulties (width : 60px, distance: 150px, and width: 80px, distance: 250px), these were empirically chosen to be challenging but completable within 2.5 seconds per target. If per iteration more than 2 out of 15 targets were missed, the iteration had to be redone. The task is shown in (\Fig{fig:task}). The utility, $y$ for the optimizer, was a weighted preference and performance score, where the performance was the per target task completion time, only hit target were taken into account. 

\begin{table}[t]
\centering
\small
\begin{tabularx}{\columnwidth}{@{} l >{\raggedright\arraybackslash}X l @{}}
\toprule
\textbf{Parameter} & \textbf{Description} & \textbf{Range} \\
\midrule
\multicolumn{3}{l}{\textbf{Software}} \\
Sensitivity & Gain from input to screen position & 0--1 \\
Reactivity & Low-pass filter (prev. pos. weight) & 0--1 \\ \addlinespace
\multicolumn{3}{l}{\textbf{Hardware}} \\
Shaft: Length & Distance from base to topper & 3--21 mm \\
Topper: Convexity & Curvature ($-$ is convex, $+$ is concave) & $\pm$0.66 \\
Topper: Width & Diameter of the topper & 10--30 mm \\
\bottomrule
\end{tabularx}
\caption{Control input parameters with descriptions and ranges. Two parameters (topper convexity and width) belong to the same component. Hardware components were prefabricated, as during the study was not feasible timewise.} 
 \Description{Software and hardware parameters with descriptions and ranges.
Two software parameters are listed: sensitivity, defined as the gain from input to screen position, ranging from 0 to 1; and reactivity, defined as a low-pass filter weighting of previous position, also ranging from 0 to 1. Three hardware parameters are listed: shaft length, the distance from base to topper, ranging from 3 to 21 millimeters; topper convexity, the curvature of the topper surface, ranging from –0.66 (convex) to +0.66 (concave); and topper width, the diameter of the topper, ranging from 10 to 30 millimeters.} 
\label{tab:userstudyparas}
\end{table}

\paragraph{Procedure}
At the start of each session, participants received standardized instructions and completed a training phase consisting of four iterations of 15 targets, each with distinct parameters. The main study proceeded iteratively: in each round, the optimizer proposed a new configuration, the joystick was reconfigured, and software parameters updated. Participants then completed four rounds of 15 Fitts's law targets. Afterward, they rated comfort using a 0–100 slider. The optimizer updated its surrogate model using the pre-specified utility weighting to select the next configuration. Sessions lasted ${\approx}75$ minutes. Joystick reconfigurations were performed by the experimenter, and their time was included in the cost model. All sessions took place in a controlled lab environment.

\begin{figure*}[t]
    \centering
    \includegraphics[width=0.8\linewidth]{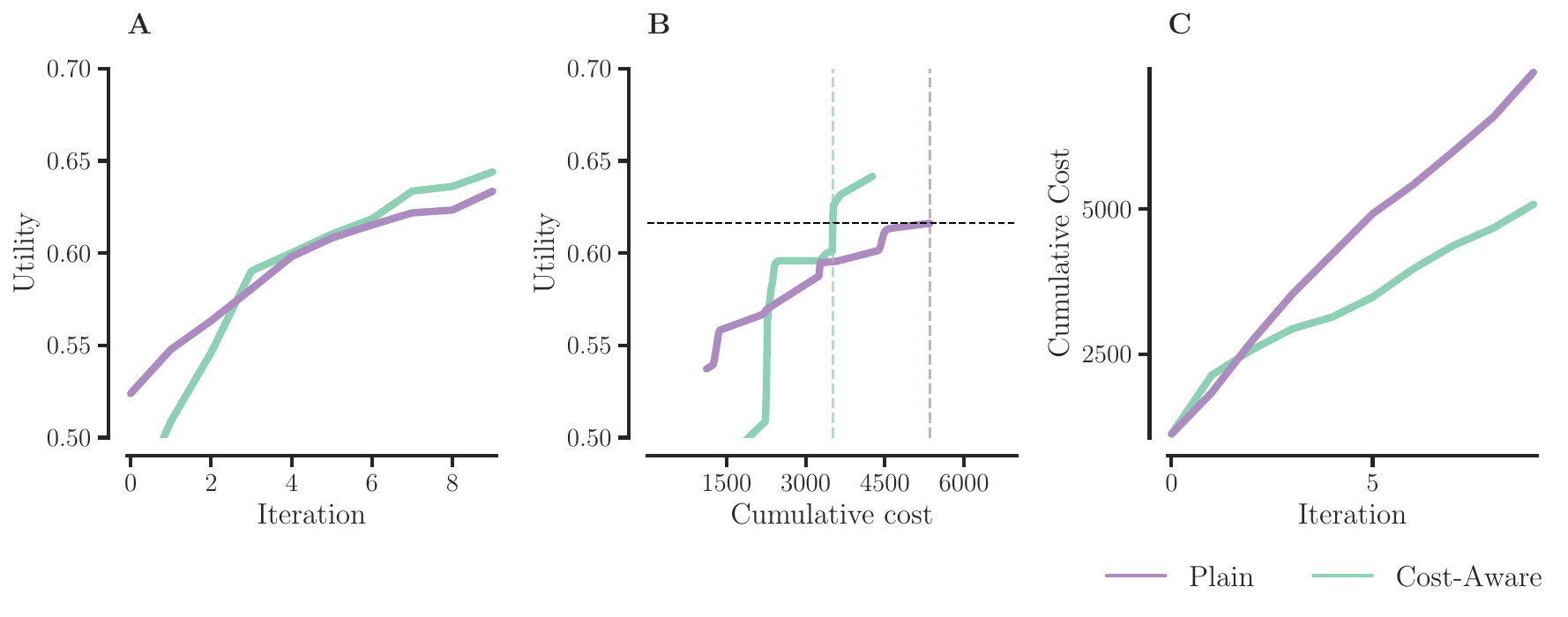}
    \caption{Results of the user study. \textbf{A}: the mean utility over the iterations. \textbf{B:} utility over cumulative cost. Our approach reaches similar quality results as the baseline in ${\approx}67\%$ of the cost. Participant dependent cumulative costs created substantial visual noise after most participants completed the task. This plot is truncated due to decreasing number of observations. \add{\textbf{C:} Our cost aware approach has a lower increase in cumulative cost compared to the baseline.}}
    \label{fig:userstudyresults}
    \Description{Utility comparison of baseline and cost-aware optimization.
Two line graphs compare utility values between baseline (purple) and cost-aware (green) methods. The left panel plots utility over iterations, showing both approaches gradually increasing. The right panel plots utility against cumulative cost, where the cost-aware method achieves comparable utility at lower cost. Dashed vertical lines mark when each method reaches the utility threshold indicated by a horizontal dashed line.
}
\end{figure*}

\paragraph{Design}
The study followed a within-subjects design, with a Latin square counterbalance. Dependent variables were the objective performance of the best configuration and participants’ subjective preferences. Both methods used Expected Improvement as the acquisition function with identical Gaussian process priors and 3 random initialization points. The optimization was done for 10 iterations. Informed consent was obtained prior to participation. No ethics approval was required under local regulations. Data were anonymized.

\subsection{Results}
The results of our study can be seen in \Fig{fig:userstudyresults}. To quantify efficiency, we analyzed the cumulative cost at which each method achieved its minimum regret. We conducted a paired-samples \texttt{t}-test to compare the number of iterations required to reach maximum utility between the \texttt{cost} and \texttt{standard} methods. The difference was statistically significant, $t(10) = 2.64$, $p = .03$, with a medium-to-large effect size ($d = 0.80$, 95\% CI $[0.10, 1.47]$). On average, the \texttt{cost} method reached maximum utility in a lower cost (\textit{M} = 3422.64, \textit{SD} = 1820.44) compared to the \texttt{standard} method (\textit{M} = 5040.46, \textit{SD} = 2034.20). \add{The final cost in our approach was lower than in the baseline (cost: \textit{M} = 5086.64, \textit{SD} = 977.33; standard: \textit{M} = 7364.36, \textit{SD} = 864.23), a difference supported by a Mann–Whitney test ($U = 6.00$, $p < .01$).} Despite the high levels of noise, due to the sample size and stochastic nature of Bayesian optimization, we can conclude that our method reaches its locally best result earlier. 

We conducted a paired-samples \textit{t}-test to compare the best result achieved by each method across trials. There was no significant difference between the \texttt{standard} method (\textit{M} = 0.37, \textit{SD} = 0.16) and the \texttt{cost} method (\textit{M} = 0.36, \textit{SD} = 0.13), $t(10) = 0.30$, $p = .77$. This is an indication that both results find an equally good configuration. 

Combined, these results show that our cost-aware method achieves outcomes comparable  to the baseline \add{at 67\% of the original cost.} Concretely, if a designer earns \$20 per hour, our method would require about \$140 compared to \$200 for the baseline. Such savings compound over multiple iterations, projects, or teams.

\paragraph{Sampling types}

\begin{figure*}[t]
    \centering
    \includegraphics[width=0.8\linewidth]{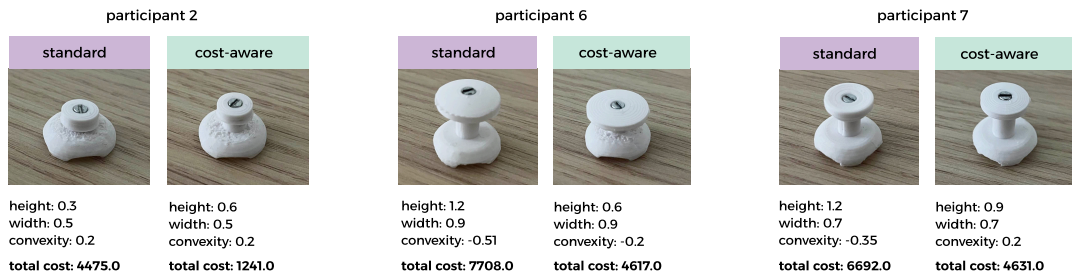}
    \vspace{1.0em}
    \caption{Qualitative examples of the best joysticks selected for three participants in both methods. Designs are similar within participants for both approaches, with the cost-aware method achieving them at lower cost. In contrast, differences appear across participants, indicating that our method supports cost-aware customization. Total cost refers to the accumulated cost of the full process}
    \label{fig:results-best}
    \Description{Example joystick prototypes produced by standard and cost-aware optimization.
Three participants’ final joystick designs are shown, with standard optimization results on the left and cost-aware results on the right. For participant 2, the cost-aware prototype doubles the height while keeping width and convexity similar, reducing cost from 4475 to 1241. For participant 6, the cost-aware version is shorter with less convexity, lowering cost from 7708 to 4617. For participant 7, the cost-aware prototype is smaller in height and width but more convex, reducing cost from 6692.0 to 4631.0.
}
\end{figure*}

Qualitative examples of the optimized joysticks suggest that the two approaches often lead to similar designs within participants, with the cost-aware method tending to do so at lower cost (\Fig{fig:results-best}). At the same time, some differences appear across participants, indicating that the method can support customization.

\section{Summary and Discussion}
We introduced a cost-aware extension to Bayesian optimization that accounts for the effort of prototyping. Across both simulations and a human-in-the-loop study, the method consistently outperformed the cost-blind baseline. In simulations, it identified designs of comparable regret at 30–40\% lower cost, adapted to structural asymmetries in the design space (e.g., modularity, dimensionality), and maintained per-iteration improvement rates despite reduced expenditure (\Sec{sec:technical}). In a user study with 12 participants customizing hardware–software joystick configurations, the method achieved equivalent task performance \add{for ${\approx}67\%$ of the original cost} (\Sec{sec:user}). These findings show that incorporating cost into the optimization loop can increase efficiency without sacrificing quality.

The core insight of our method is that cost in HCI prototyping is not fixed but depends on design history. What matters is not only a configuration’s parameter values, but also what has already been built, what can be reused, and how parameters map onto physical or software components. Our cost model makes this dependency explicit through three reuse categories applied at the component level, with a prototype record capturing past builds and their recency. This formulation makes cost-aware optimization both feasible and meaningful in design settings where effort varies widely. Importantly, it does so without altering how utility is defined or measured, only how effort is accounted for.

To apply the cost model in a prototyping workflow, researchers define components, assign reuse categories, and integrate these into the optimization loop. The process consists of five steps: \textbf{(1)} group design parameters into components that reflect how the system is built and modified (e.g., enclosures, sensors, or control filters); \textbf{(2)} assign each component a reuse category—\tweak if unchanged from the last iteration, \swap if reused from earlier prototypes, or \create if newly fabricated or implemented; \textbf{(3)} map each category to a scalar cost, either using coarse values (e.g., 1/10/100) or estimates based on effort, with consistency being more important than precision; \textbf{(4)} compute the configuration cost during optimization and incorporate it into the acquisition function, guiding the optimizer toward cost-effective, high-impact edits; and \textbf{(5)} update the cost structure as iteration progresses, so the model adapts as fabrication methods evolve or cost priorities change. This workflow is lightweight and adaptable, making reuse explicit and cost-aware.

Ultimately, this work enables more effective use of optimization in prototyping scenarios where cost is a central concern. Our method shifts the focus from sample efficiency to cost efficiency, aligning the exploration–exploitation balance of the optimizer with the considerations designers face when deciding what to prototype.

\subsection{Limitations and Future Work}
Our current model assumes that components change independently. In practice, many physical and software edits are interdependent: replacing a sensor may require redesigning a housing, adjusting power routing, and updating software calibration~\cite{riesener2019model}. A more structured cost model, such as a tree representation, could better capture these effects.

Cost also varies with fidelity. Designers frequently use low-fidelity prints, mockups, or simulations before committing to high-fidelity builds~\cite{song2019general}. Our method assumes uniform evaluation fidelity, yet fidelity itself is a design decision. Extending the approach with multi-fidelity Bayesian optimization~\cite{do2023multi} could help balance confidence against expense, supporting broader early-stage exploration while deferring costlier evaluations.

Our method operates entirely at the acquisition-function level, which allows broad compatibility with different forms of feedback. As long as the surrogate model can incorporate it, the acquisition function can use objective scores, subjective ratings, preferences, or rejections. This makes the approach suitable for human-in-the-loop workflows where designers steer the process, exclude infeasible samples, or adjust priorities~\cite{yamamoto2022photographic}.

Acquisition functions do more than select samples --- they define the purpose of optimization. Our method assumes the goal is to maximize utility quickly, through maximizing expected improvement. But in early-stage prototyping, the goal may be understanding rather than performance. Acquisition strategies that target information gain or uncertainty reduction~\cite{noe2018new} could better support these phases. Future tools might dynamically shift acquisition behavior over time, blending exploration and exploitation in response to available budget, designer input, or changing goals~\cite{fisch2012techniques}.

\add{A further direction is to extend the cost model beyond a single metric. Some costs cannot be mapped to a common scale and may be better handled as separate budget dimensions. A multi-budget formulation would let designers balance heterogeneous resources and choose configurations along a Pareto front. Integrating such a model into Bayesian optimization and understanding how designers work with these trade-offs are promising areas for future work.}

\add{Perhaps the most critical question for all applications of Bayesian optimization in design is that it requires a predefined parametric design space. Advancing methods that can infer, expand, or adapt the design space during the optimization process would make these approaches more viable for real prototyping workflows.}

\section{Conclusion} 
Cost is central to prototyping: it determines what gets built, how often teams iterate, and which directions are feasible at all. We \addcr{adapted} cost-aware optimization to support these decisions. It builds on two ideas: classifying changes into structured cost types (\tweak, \swap, and \create), and maintaining a \emph{prototype record} that tracks which components have already been built. Together, these form a model of cost that reflects real prototyping effort. By integrating this model into the acquisition function, our method selects configurations based on expected improvement per unit cost, encouraging reuse and avoiding rebuilds. It fits into existing Bayesian optimization frameworks, requiring no changes to the surrogate model. Simulations and a user study show that it reduces cost without degrading performance. More broadly, our work reframes optimization in HCI as a resource allocation problem: not just finding high-performing designs, but deciding what is worth building next under real budget constraints.

\begin{acks}
This work was funded by the European Research Council (101141916) and the Research Council of Finland (328400, 345604, 341763, and 357578). The code is freely available: \url{https://github.com/aalto-ui/CABOP}.
\end{acks}

\bibliographystyle{ACM-Reference-Format}
\bibliography{references}

\clearpage
\appendix
\section{Application Examples}
\label{app:cases}

\subsubsection*{Example 1: Dexmo — a mechanical exoskeleton for motion capture and force feedback in VR}

\paragraph{Purpose} Dexmo is a lightweight mechanical exoskeleton designed for finger-level motion tracking and passive force feedback in virtual reality \cite{gu2016dexmo}. It uses servo-actuated joint locking to simulate rigid contacts without requiring large actuators or tethers.

\paragraph{Components} Dexmo can be decomposed into three primary components: (1) the finger linkage mechanism, (2) the force feedback unit (FFU), and (3) the controller and communication module. Each of these has tunable parameters. The linkage varies in rod length and joint placement. The FFU includes servo type and locking geometry. The controller allows changes to sampling rate, filtering strategy, and resolution.

\paragraph{Cost Structure} Reuse is straightforward. Previously printed linkage parts can often be swapped in new configurations (\swap), while software changes like filtering are typically tweaks (\tweak). Replacing or redesigning the FFU involves high precision and servo calibration, making it a costly \create operation. This structure allows the optimizer to reason about cost across both hardware and software components. How to obtain costs?

\paragraph{Optimization Objective} The system could be optimized for user-perceived realism or quantitative tracking accuracy. Our cost model allows the system to explore whether performance gains from a new FFU justify fabrication cost, while encouraging reuse of components that do not need to change.

\subsubsection*{Example 2: Omni — an electromagnetic haptic feedback device}

\paragraph{Purpose} Omni is a volumetric haptic device that combines input and output through a stylus and custom electromagnetic actuator \cite{langerak2020omni, zarate2020contact}. A permanent magnet embedded in the stylus interacts with a spherical electromagnet, while a Hall sensor array tracks stylus position.

\paragraph{Components} The device can be decomposed into three components: the stylus, the electromagnet, and the sensor array. The electromagnet has parameters for core size, winding count, and wire gauge, often requiring finite element simulation. The stylus can be modified by changing magnet length and radius. The sensor array varies in number, layout, and filter settings.

\paragraph{Cost Structure} The electromagnet is costly to change (\create), involving modeling, winding, and high material costs. The stylus and sensor configurations can often be reused (\swap), and some software-level changes such as filter length are minor (\tweak). These component-level distinctions allow the optimizer to isolate expensive modifications and favor reuse where possible.

\paragraph{Optimization Objective} Omni could be optimized through human ratings of haptic fidelity or through measured force and tracking accuracy. The cost model would help manage effort by minimizing unnecessary hardware changes and prioritizing high-impact tuning steps.

\vspace{3em}

\subsubsection*{Example 3: Back-Hand-Pose — wrist-mounted camera for hand pose estimation using machine learning}

\paragraph{Purpose} Back-Hand-Pose is a real-time system for 3D hand pose estimation using a wrist-mounted RGB camera facing the back of the hand \cite{wu2020back}. It uses a two-stream LSTM-based network (DorsalNet) to infer hand joint positions from skin deformations, without relying on external cameras or markers.

\paragraph{Components} The system consists of three components: the wrist-mounted camera, the preprocessing pipeline (e.g., masking and motion history), and the neural network model. Parameters include camera field of view, placement, image transformations, and model architecture.

\paragraph{Cost Structure} The dominant cost in this system is retraining. Changes in camera placement or data representation can invalidate prior training data, triggering expensive data collection and training cycles (\create). Changes to the preprocessing logic may allow reuse of data and partial retraining (\swap). Tuning hyperparameters such as learning rate is typically inexpensive in terms of human effort but may incur wall-time costs (\tweak). This structure highlights how software prototyping, like hardware, benefits from a cost-sensitive view.

\paragraph{Optimization Objective} The system can be optimized for pose accuracy, robustness, or generalization. A cost-aware optimizer can balance retraining cost against model performance, allowing selective improvements without restarting the entire pipeline.

\begin{figure}[h]
    \centering
    \begin{subfigure}[b]{0.45\linewidth}
        \centering
        \includegraphics[width=\linewidth]{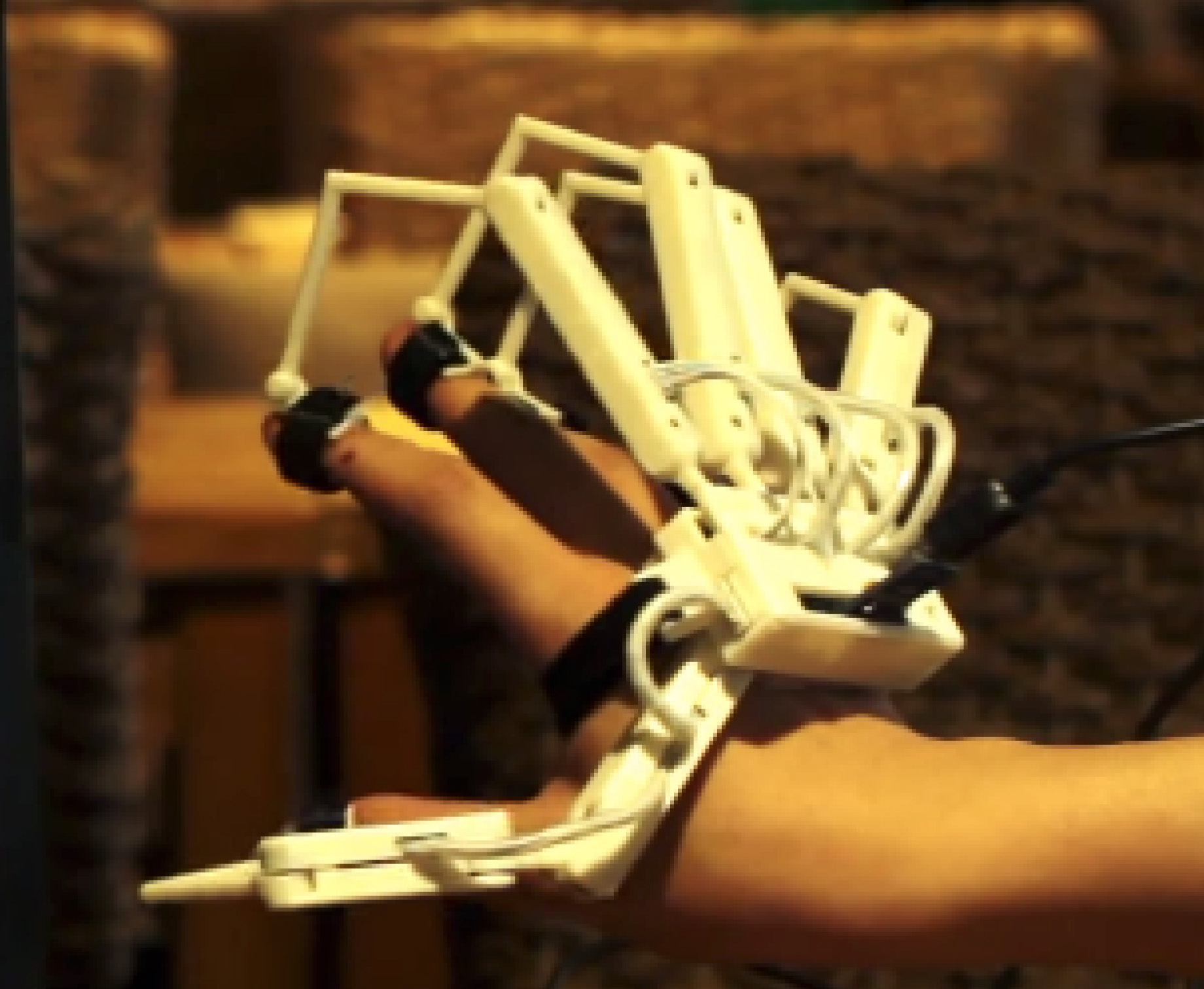}
        \caption{Dexmo \cite{gu2016dexmo}}
        \label{fig:dexmo}
    \end{subfigure}
    \begin{subfigure}[b]{0.45\linewidth}
        \centering
        \includegraphics[width=\linewidth]{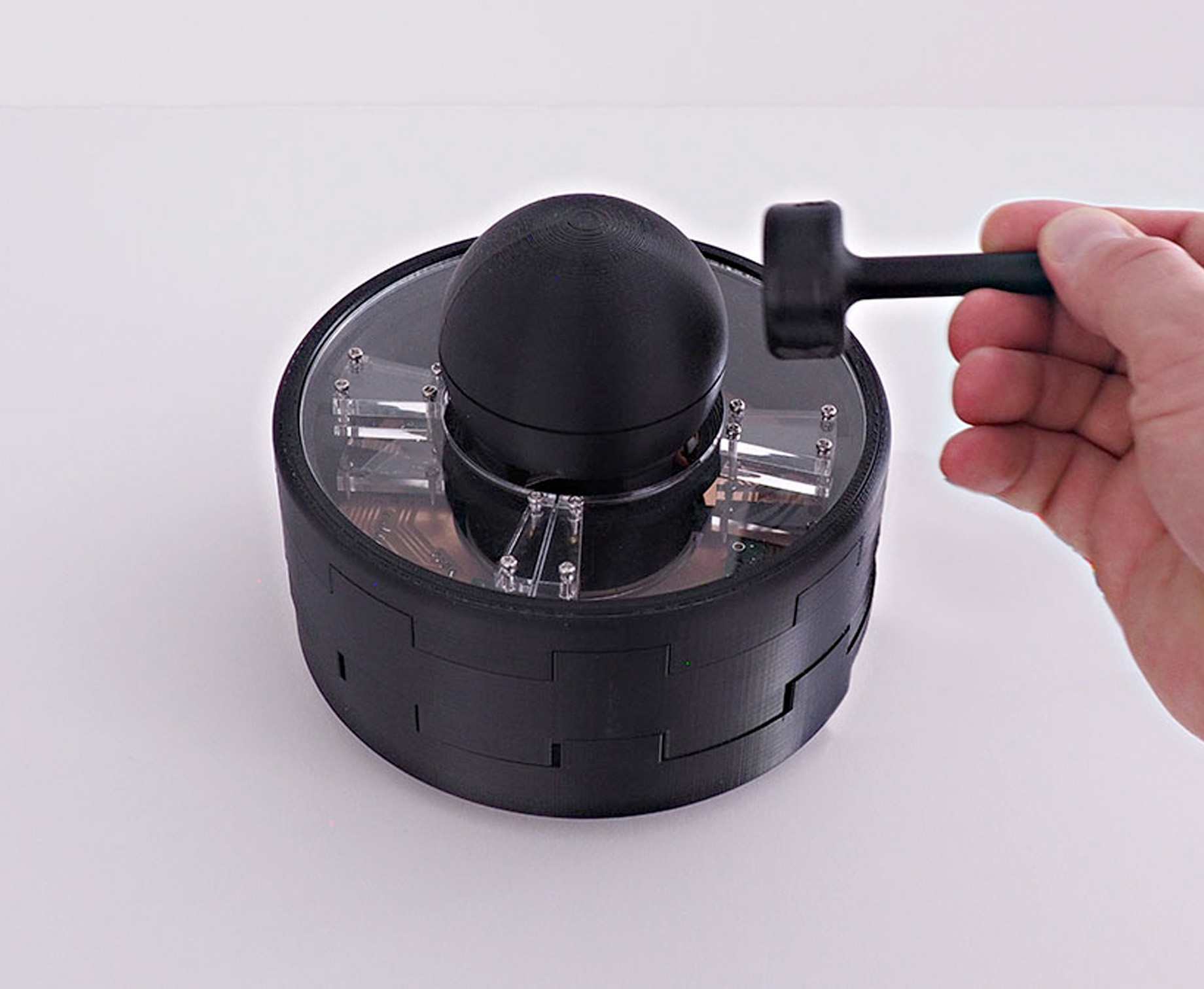}
        \caption{Omni \cite{langerak2020omni, zarate2020contact}}
        \label{fig:omni}
    \end{subfigure}
    \begin{subfigure}[b]{0.45\linewidth}
        \centering
        \includegraphics[width=\linewidth]{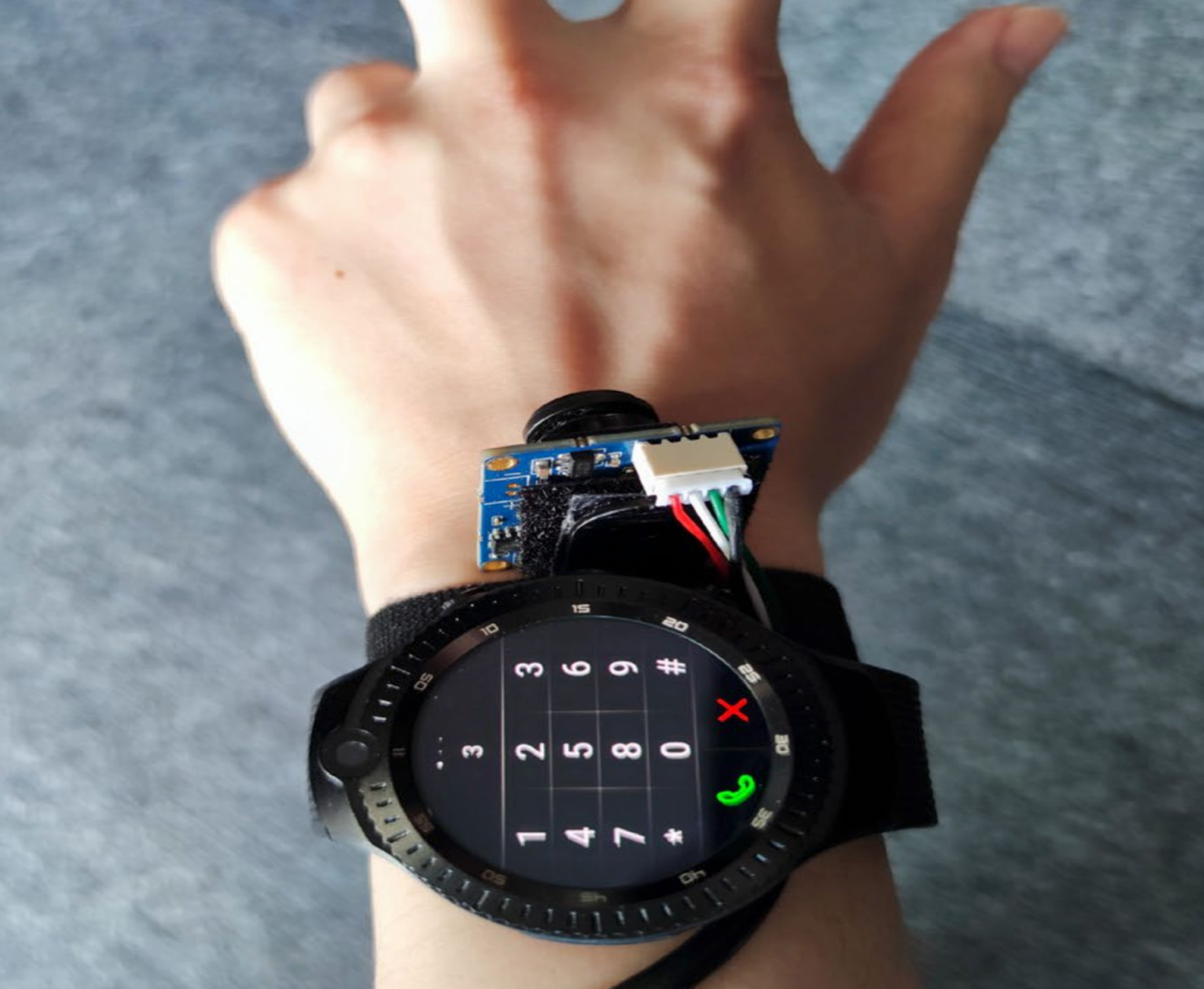}
        \caption{Back-Hand-Pose \cite{wu2020back}}
        \label{fig:backhand}
    \end{subfigure}
    \caption{Retrospective application of our cost model to three systems from the HCI literature. Each system features distinct component structures and cost asymmetries, demonstrating the model's flexibility across domains.}
    \Description{Examples of interactive devices from prior work.
Panel (a) shows Dexmo, a mechanical exoskeleton glove worn on the hand. Panel (b) shows Omni, a black spherical joystick device mounted on a base and controlled with a handle. Panel (c) shows Back-Hand-Pose, a wrist-worn smartwatch prototype with added sensors and an on-screen numeric interface for hand pose recognition.
}
    \label{fig:exampleapps}
\end{figure}

\clearpage
\section{User Study}
\label{app:user}

\begin{figure}[h]
    \centering
    \includegraphics[width=\linewidth]{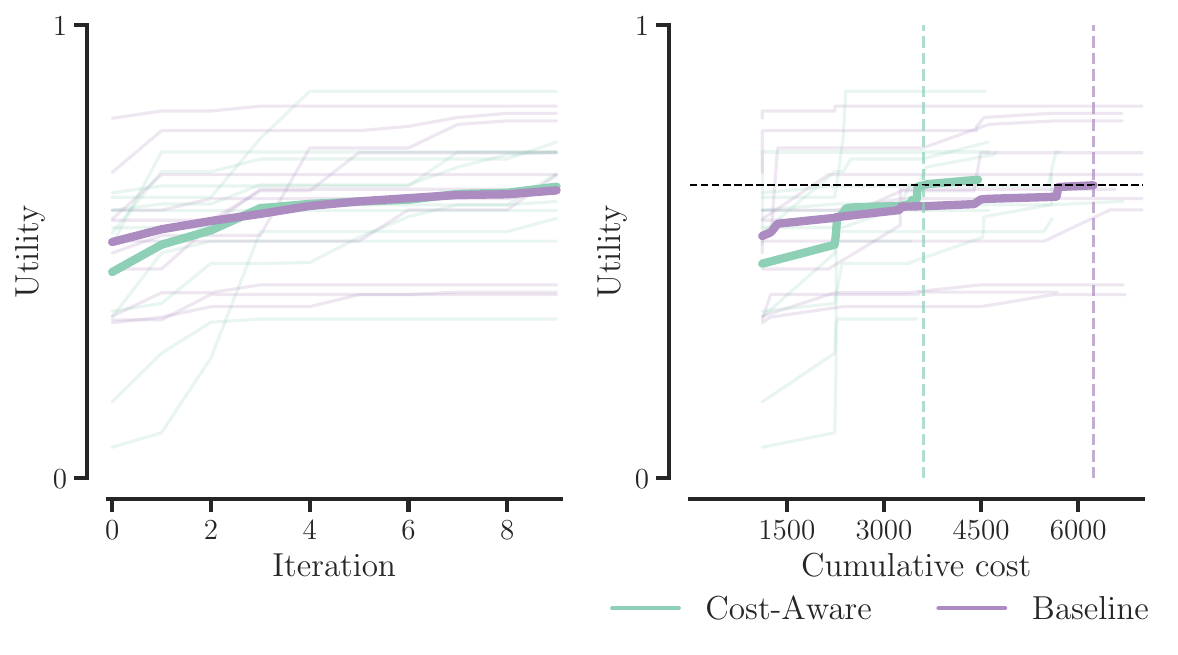}
    \caption{Results of the user study, both plots show individual users in the back ground. \textbf{Left}: the utility over the iterations. \textbf{Right:} utility over cumulative cost. Our approach reaches similar quality results as the baseline in $\sim70\%$ of the cost.}
    \label{fig:userstudyresults_appendix}
    \Description{Utility comparison of baseline and cost-aware optimization.
Two line graphs compare utility values between baseline (purple) and cost-aware (green) methods. The left panel plots utility over iterations, showing both approaches gradually increasing. The right panel plots utility against cumulative cost, where the cost-aware method achieves comparable utility at lower cost. Dashed vertical lines mark when each method reaches the utility threshold indicated by a horizontal dashed line. 
}
\end{figure}

\add{
    \section{Additional Ground Truth Functions}
\label{app:extragt}

In the main paper (\Sec{sec:technical}), we benchmarked against a Bayesian optimization baseline commonly used in HCI~\cite{liao2023human}, which selects candidates via expected improvement without considering cost. Our approach instead optimizes expected improvement normalized by cost. For systematic evaluation, we used the 2D Rosenbrock function as the ground-truth utility surface. 

To evaluate our approach and remove the effect of the ground-truth function, we report results on three additional functions, beyond the Rosenbrock: Ackley~\cite{ackley2012connectionist}, Goldstein-Price~\cite{picheny2013benchmark}, and Levy~\cite{laguna2005experimental}. These functions exhibit distinct behaviors, demonstrating that our method generalizes beyond a single test function. 

\paragraph{Ackley.}
The Ackley function is widely used to test optimization methods due to its many local minima surrounding a global minimum at the origin:
\begin{equation}
\begin{aligned}
    f(x_1, x_2)  =  & -20 \exp\left(-0.2 \sqrt{0.5(x_1^2 + x_2^2)}\right) - \exp\bigg(0.5 (\cos(2 \pi x_1) \\ 
    & + \cos(2 \pi x_2))\bigg) + 20 + e
\end{aligned}
\end{equation}
In HCI prototyping terms, it reflects design problems where many suboptimal solutions appear promising (local optima), yet only careful exploration reveals the true global solution.

\paragraph{Goldstein–Price.}
The Goldstein–Price function has a complex surface with multiple local minima and a single global minimum at $(0, -1)$:
\begin{equation}
\begin{aligned}
    &f(x_1, x_2) = \\
    &\quad \big[1 + (x_1 + x_2 + 1)^2 (19 - 14x_1 + 3x_1^2 - 14x_2 + 6x_1x_2 + 3x_2^2)\big] \times \\
    &\quad \big[30 + (2x_1 - 3x_2)^2 (18 - 32x_1 + 12x_1^2 + 48x_2 - 36x_1x_2 + 27x_2^2)\big]
\end{aligned}
\end{equation}
Its multiple basins of attraction mimic design situations with several viable prototypes that compete in quality, requiring strategies that distinguish globally superior solutions from merely adequate ones.

\paragraph{Levy.}
The Levy function is multimodal, with its global minimum at $(1,1)$. Its definition is:
\begin{equation}
\begin{aligned}
    f(x_1, x_2) &= \sin^2(\pi w_1) + (w_1 - 1)^2 \big[1 + 10 \sin^2(\pi w_1 + 1)\big] \\
    &\quad + (w_2 - 1)^2 \big[1 + \sin^2(2 \pi w_2)\big], \\
    w_i &= 1 + \frac{x_i - 1}{4}, \quad i \in \{1,2\}
\end{aligned}
\end{equation}
For HCI, it captures scenarios where progress requires navigating highly irregular design spaces, where small parameter changes can lead to large variations in outcome quality, as in tuning interactive hardware or software parameters under uncertainty.

\paragraph{Validity}
For each experiment we will only provide a brief introduction, and refer to the main paper for more information. The evaluations are run exactly as reported in the main paper, and only the ground truth function changes. We focus on additional results for all studies, except study 4 as it requires a higher-dimensional formulation and is therefore omitted here. 

All studies rely on multiple randomized instances, multiple repetitions, and random seeds per repetition. Known optima allow objective evaluation, ensuring results are robust, interpretable, and reproducible.

\begin{figure*}[ht!]
    \centering

    \begin{subfigure}{0.47\linewidth}
        \centering
        \includegraphics[width=\linewidth]{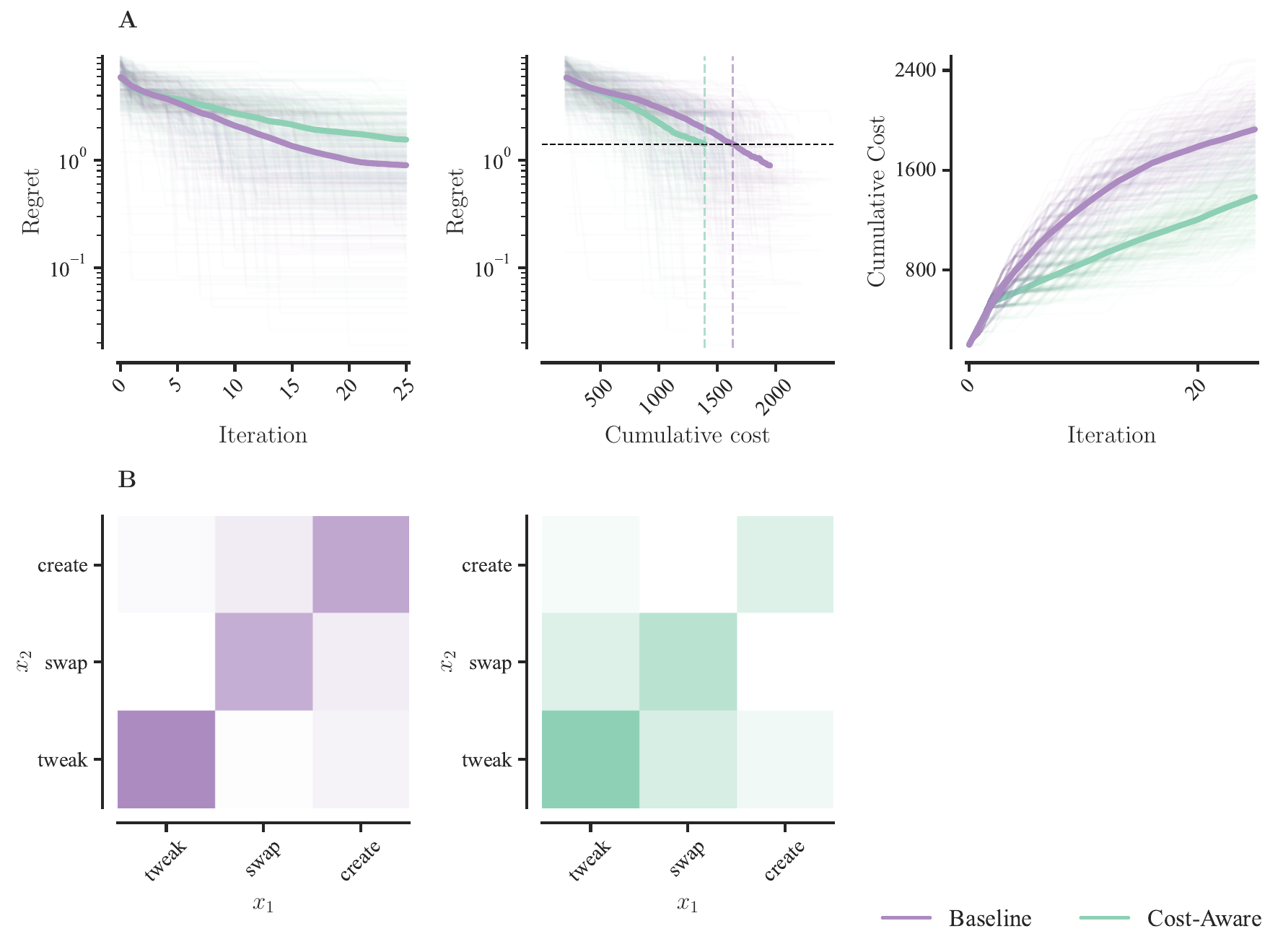}
        \subcaption{Ackley}
        \label{fig:ackley_study1}
    \end{subfigure}
    \hfill
    \begin{subfigure}{0.47\linewidth}
        \centering
        \includegraphics[width=\linewidth]{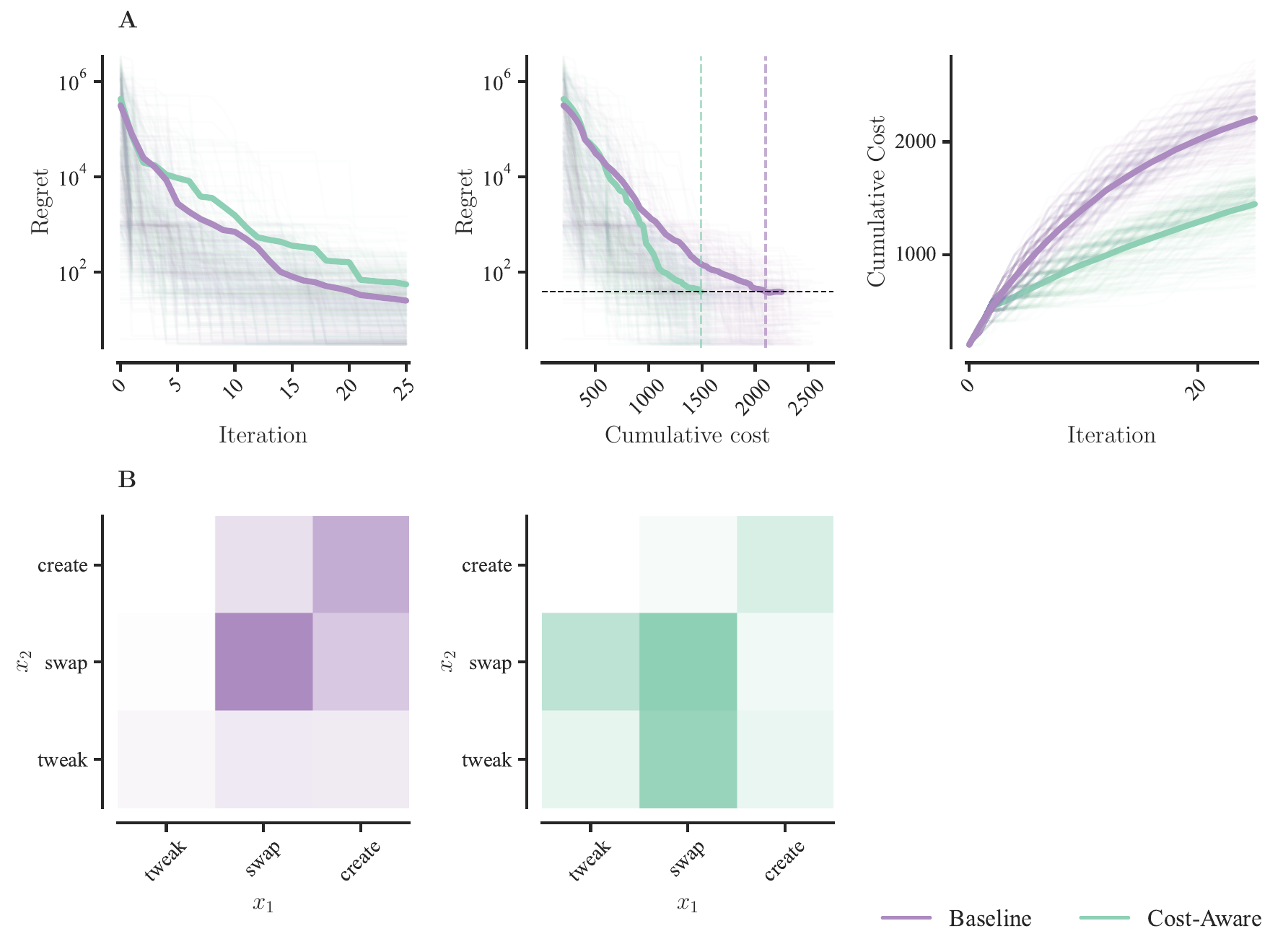}
        \subcaption{Goldstein--Price}
        \label{fig:goldstein_study1}
    \end{subfigure}

    \vspace{1em}

    \begin{minipage}[t]{0.47\linewidth}
        \centering
        \includegraphics[width=\linewidth]{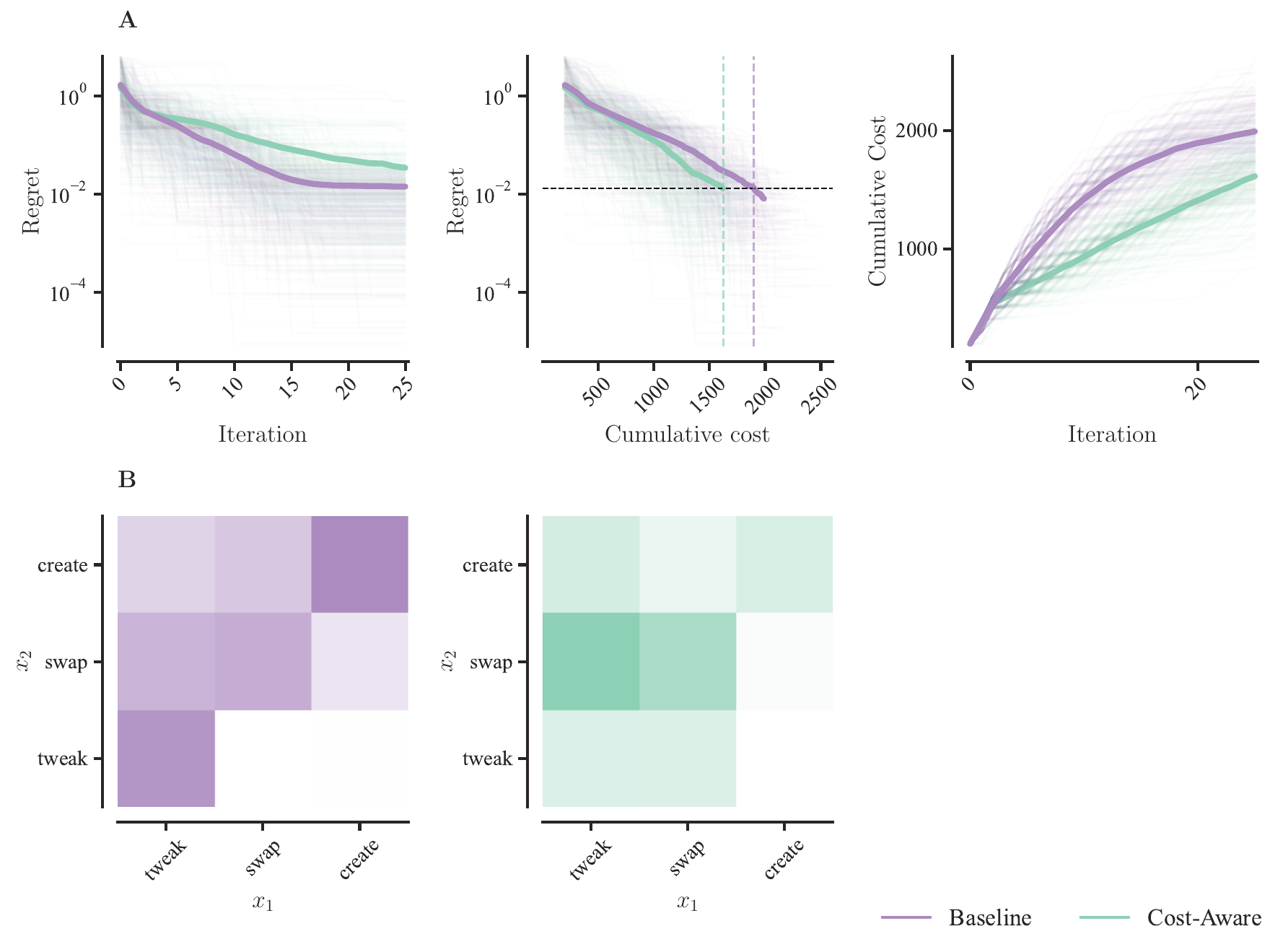}
        \subcaption{Levy}
        \label{fig:levy_study1}
    \end{minipage}
    \hfill
    \begin{minipage}[t]{0.47\linewidth}
        \vspace*{-2.75cm} %
        \captionof{figure}{
            \add{\textbf{Study 1:} Comparison of cost-aware and baseline Bayesian optimization under a fixed number of iterations.
            $x_1$ and $x_2$ are two parameters in $\design$.  
            \textbf{A - Left:} Regret over iterations.  
            \textbf{A - Center:} Regret versus cumulative cost; the cost-aware method achieves similar regret at lower cost.  
            \textbf{A - Right:} Cumulative cost across iterations; cost-aware consistently requires less cost.  
            \textbf{B:} Distribution of sampled edit types (\tweak, \swap, \create); cost-aware sampling is more balanced.}
        }
        \Description{Comparison of regret, cost, and sampling behavior across benchmark functions for baseline and cost-aware optimization.
The figure presents results for three objective functions: Ackley, Goldstein–Price, and Levy.
For each function, Panel A shows three line plots: regret over iterations, regret versus cumulative cost, and cumulative cost over iterations. In all cases, regret decreases for both methods, and the cost-aware method reaches similar regret at lower cumulative cost.
Panel B shows heatmaps of sampling frequencies across tweak, swap, and create actions. Baseline sampling concentrates in high-cost create actions, whereas the cost-aware method shifts sampling toward lower-cost actions, with patterns varying by function.

        }
    \end{minipage}
\end{figure*}

\subsection{Study 1: Performance Under Unlimited Budget}
This study tests whether the cost-aware method achieves similar outcomes with less cumulative cost, even when no explicit budget is imposed. We optimized each groundtruth function over 25 iterations with 250 independent trials per method. Both methods began with three random Sobol samples. At each step, we recorded the regret and the action cost type (\tweak, \swap, \create). While iteration count was fixed, cumulative cost was not, allowing us to compare sample efficiency (steps) and cost efficiency (resources used). Statistical analysis used Mann–Whitney U tests due to non-normality of distributions.

\subsubsection*{Ackley.}
Results are shown in Fig~\ref{fig:ackley_study1}. At the point of lowest regret, the cost-aware method required significantly less cost ($M = 1010.23$, $SD = 302.72$) than the baseline ($M = 1448.25$, $SD = 361.34$), Mann–Whitney $U = 10504.00$, $p < .001$. However, final regret was higher for cost-aware ($M = 1.56$, $SD = 1.05$) than baseline ($M = 0.90$, $SD = 0.68$), $U = 43808.00$, $p < .001$. At a low-cost budget, the lowest total cost out of all runs, ($\text{budget} = 781$), cost-aware achieved lower regret ($M = 3.21$, $SD = 1.24$) than baseline ($M = 3.98$, $SD = 1.36$), $U = 20582.00$, $p < .001$. Using the complete sample, the final cost in our approach was lower than in the baseline (cost: \textit{M} = 1386.23, \textit{SD} = 248.49; standard: \textit{M} = 1927.28, \textit{SD} = 227.48), a difference confirmed by a two-sample t-test ($t(498) = -25.39$, $p < .001$). Neither method reached the global optimum. These results suggest that while cost-aware incurs slightly higher final regret, it consistently does so at substantially lower cost.

\subsubsection*{Goldstein–Price.}
Results are shown in Fig~\ref{fig:goldstein_study1}. At the point of lowest regret, the cost-aware method required significantly less cost ($M = 998.80$, $SD = 312.89$) than the baseline ($M = 1590.56$, $SD = 514.59$), $U = 10731.50$, $p < .001$. Final regret was again higher for cost-aware ($M = 55.60$, $SD = 116.68$) than for baseline ($M = 25.32$, $SD = 33.97$), $U = 39169.00$, $p < .001$. At a matched cost level, the lowest final cost of all runs, ($\text{budget} = 709$), cost-aware achieved substantially lower regret ($M = 8973.39$, $SD = 51980.67$) than the baseline ($M = 15437.55$, $SD = 60678.71$), $U = 22441.00$, $p < .001$. Using the full sample, the final cost in our approach was lower than in the baseline (cost: \textit{M} = 1447.29, \textit{SD} = 206.94; standard: \textit{M} = 2207.46, \textit{SD} = 281.08), a difference supported by a Mann–Whitney test ($U = 1781.00$, $p < .001$).Thus, although final regret is higher, cost-aware consistently reduces regret at equivalent or lower cost levels.

\subsubsection*{Levy.}
Results are shown in Fig~\ref{fig:levy_study1}. At the point of lowest regret, cost-aware required significantly less cost ($M = 1164.48$, $SD = 318.99$) than baseline ($M = 1614.22$, $SD = 370.13$), $U = 9672.50$, $p < .001$. Final regret was higher for cost-aware ($M = 0.04$, $SD = 0.09$) than for baseline ($M = 0.01$, $SD = 0.06$), $U = 39687.00$, $p < .001$. At a comparable cost level ($\text{budget} = 835$), the difference remained significant, $U = 26461.00$, $p = .003$. Using all samples, the final cost in our approach remained lower than in the baseline (cost: \textit{M} = 1616.56, \textit{SD} = 233.88; standard: \textit{M} = 1995.60, \textit{SD} = 240.01), a difference supported by a Mann–Whitney test ($U = 7403.00$, $p < .001$). These findings indicate that while final regret is higher, cost-aware consistently requires less cost to achieve comparable outcomes.

\subsection{Study 2: Performance Under Budget Constraints}
This study evaluates whether the cost-aware method achieves better results when the total available budget is capped. We simulated optimization under fixed cost budgets, the range was determined on a per function basis. Each method used its entire budget and terminated once exhausted. For each budget level, we recorded the final regret. At each level, 75 optimization runs were performed per method. As in Study 1, we used Mann–Whitney U tests to assess statistical significance due to non-normal outcome distributions.

\begin{figure*}[ht]
    \centering

    \begin{subfigure}{0.8\linewidth}
        \centering
        \includegraphics[width=\linewidth]{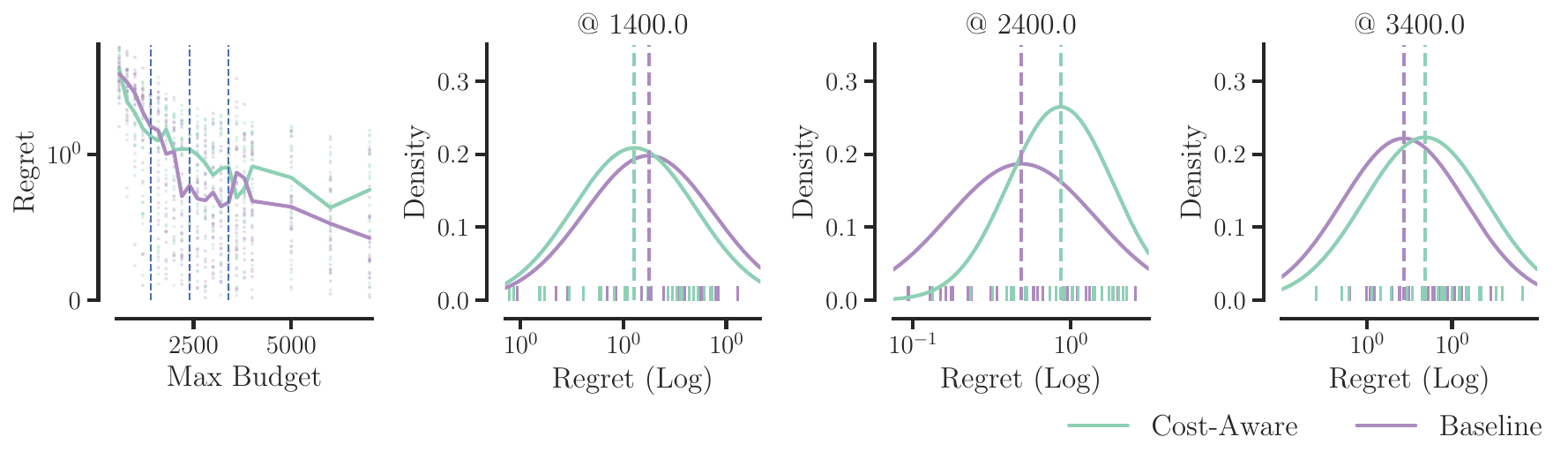}
        \subcaption{Ackley %
        }
        \label{fig:ackley_study2}
    \end{subfigure}

    \vspace{1em}

    \begin{subfigure}{0.8\linewidth}
        \centering
        \includegraphics[width=\linewidth]{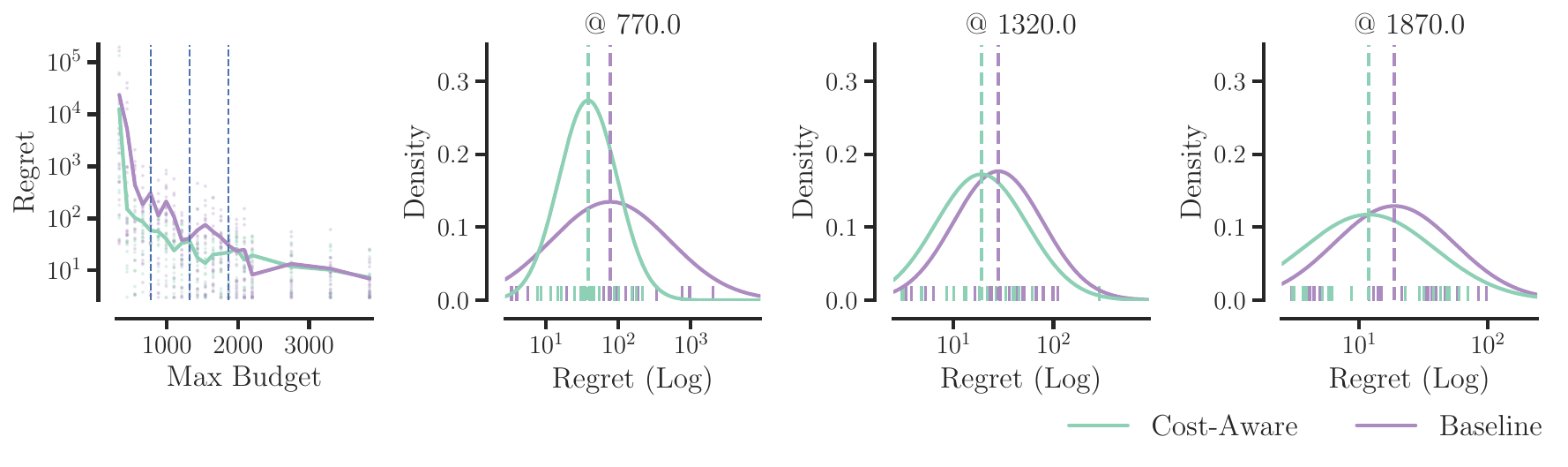}
        \subcaption{Goldstein–Price %
        }
        \label{fig:goldstein_study2}
    \end{subfigure}

    \vspace{1em}

    \begin{subfigure}{0.8\linewidth}
        \centering
        \includegraphics[width=\linewidth]{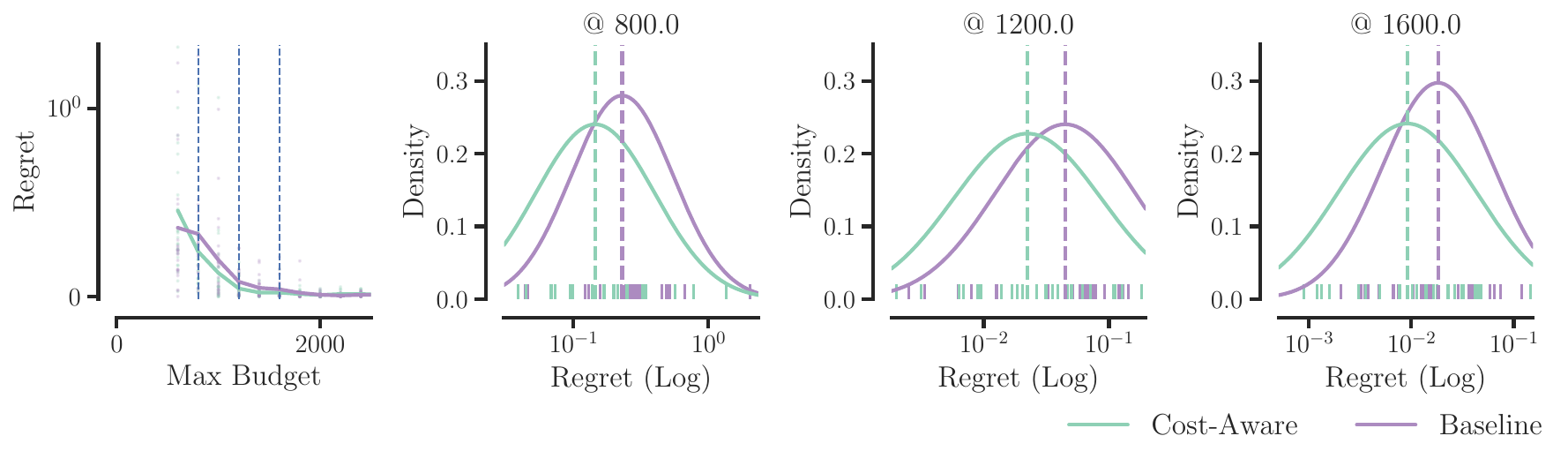}
        \subcaption{Levy %
        }
        \label{fig:levy_study2}
    \end{subfigure}

    \caption{
        \textbf{Study 2:} Final Regret and Regret Distributions Across Cost Budgets. Each row shows a benchmark function. 
        \textbf{Left:} Final regret across a range of cost budgets, with distributions per budget. 
        \textbf{Right:} Regret distributions at selected cost budgets. 
        The cost-aware method tends to outperform the baseline at low and mid-range budgets, with differences varying across the three functions.
    }
    \Description{Regret trends and regret distributions across benchmark functions for baseline and cost-aware optimization.
The figure shows results for three objective functions: Ackley, Goldstein–Price, and Levy.
For each function, the first plot shows regret decreasing as the maximum budget increases, with the cost-aware method reaching lower regret earlier.
The following three density plots show the distribution of regret at selected budget levels. In each case, the cost-aware distribution shifts toward lower regret compared with the baseline, indicated by shifted peaks and vertical markers.

    }
    \label{fig:study2_combined}
\end{figure*}

\subsubsection*{Ackley.}
Figure~\ref{fig:ackley_study2} shows that at a budget of 1400 the cost-aware method achieved slightly lower regret than the baseline, but the difference was not significant ($p = .065$). At higher budgets (2400 and 3400), the baseline performed significantly better ($p = .001$ and $p = .027$). Thus, only the lowest budget weakly favored the cost-aware method. This limited effect likely reflects the Ackley landscape: evenly distributed local minima and smooth transitions reduce sensitivity to cost allocation, so both methods converge to comparable solutions. This is also reflected by the noise shown in the left figure, where even higher budgets are not guarantee to finding a better solution for either approach. 

\subsubsection*{Goldstein–Price.}
Results are shown in Fig.~\ref{fig:goldstein_study2}. At a budget of 770, regret was significantly lower for the cost-aware method ($M = 58.58$, $SD = 58.20$) than for the baseline ($M = 304.76$, $SD = 496.81$), $U = 1296.00$, $p = .008$. At 1320, the cost-aware method again outperformed the baseline ($M = 36.01$, $SD = 60.76$ vs.\ $M = 41.51$, $SD = 30.20$), $U = 1233.00$, $p = .003$. At 1870, the cost-aware method maintained lower regret ($M = 22.20$, $SD = 22.01$) compared to the baseline ($M = 30.09$, $SD = 26.25$), but this difference was not statistically significant ($U = 1485.00$, $p = .099$). These findings indicate consistent advantages for the cost-aware method at lower and mid-range budgets, though the gap narrows at higher allocations.

\subsubsection*{Levy.}
Results are shown in Fig.~\ref{fig:levy_study2}. At a budget of 800, regret was significantly lower for the cost-aware method ($M = 0.24$, $SD = 0.29$) compared to the baseline ($M = 0.33$, $SD = 0.38$), $U = 1944.00$, $p = .001$. At 1200, this difference remained, with the cost-aware method again showing lower regret ($M = 0.04$, $SD = 0.05$) than the baseline ($M = 0.08$, $SD = 0.07$), $U = 1881.00$, $p < .001$. At 1600, the methods continued to diverge, with the cost-aware method ($M = 0.02$, $SD = 0.03$) outperforming the baseline ($M = 0.04$, $SD = 0.05$), $U = 2259.00$, $p = .038$. These results demonstrate consistent and statistically reliable advantages for the cost-aware method across budgets, with particularly strong improvements visible at lower and mid-range budgets.

\begin{figure*}[ht]
    \centering
    \begin{subfigure}{0.75\linewidth}
        \centering
        \includegraphics[width=\linewidth]{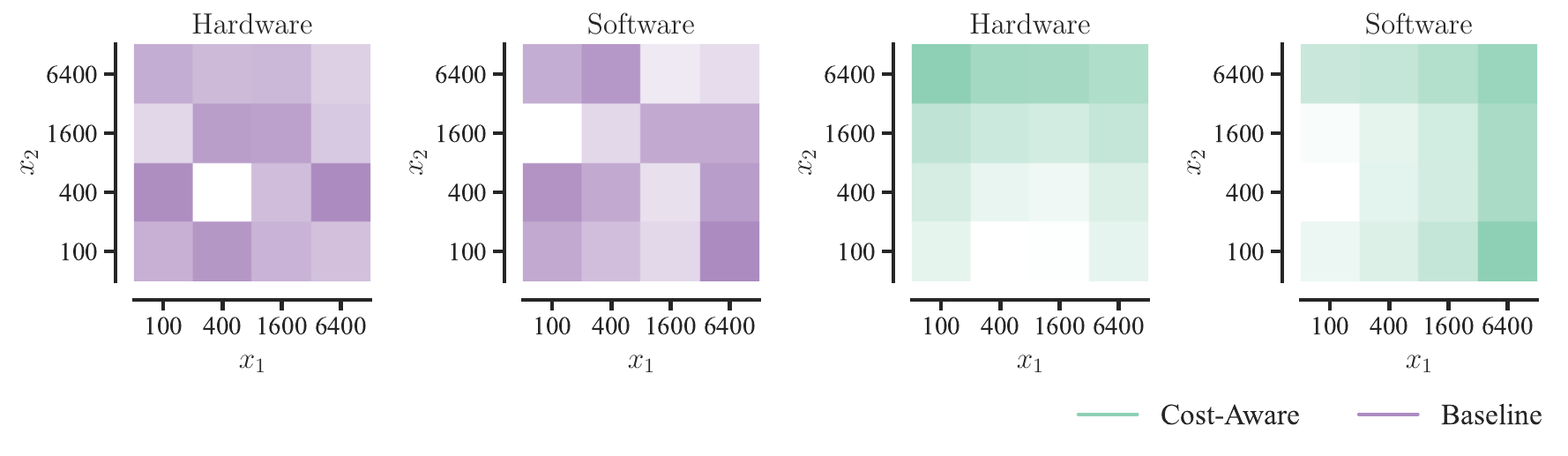}
        \subcaption{Ackley %
        }
        \label{fig:ackley_study3}
    \end{subfigure}

    \vspace{1.2em}

    \begin{subfigure}{0.75\linewidth}
        \centering
        \includegraphics[width=\linewidth]{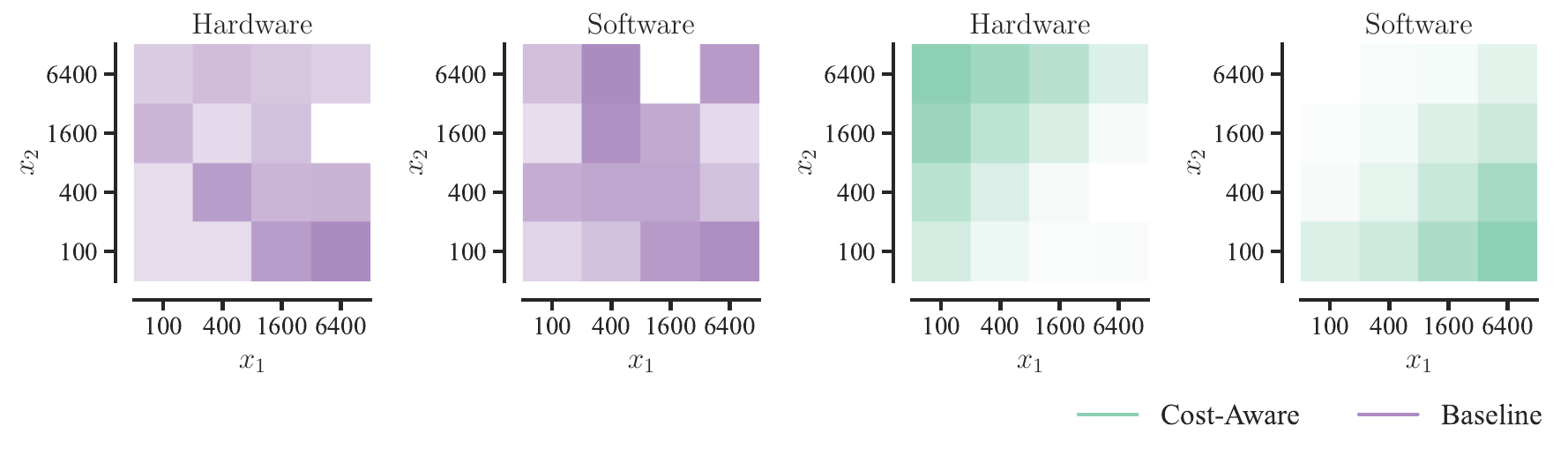}
        \subcaption{Goldstein–Price %
        }
        \label{fig:goldstein_study3}
    \end{subfigure}

    \vspace{1.2em}

    \begin{subfigure}{0.75\linewidth}
        \centering
        \includegraphics[width=\linewidth]{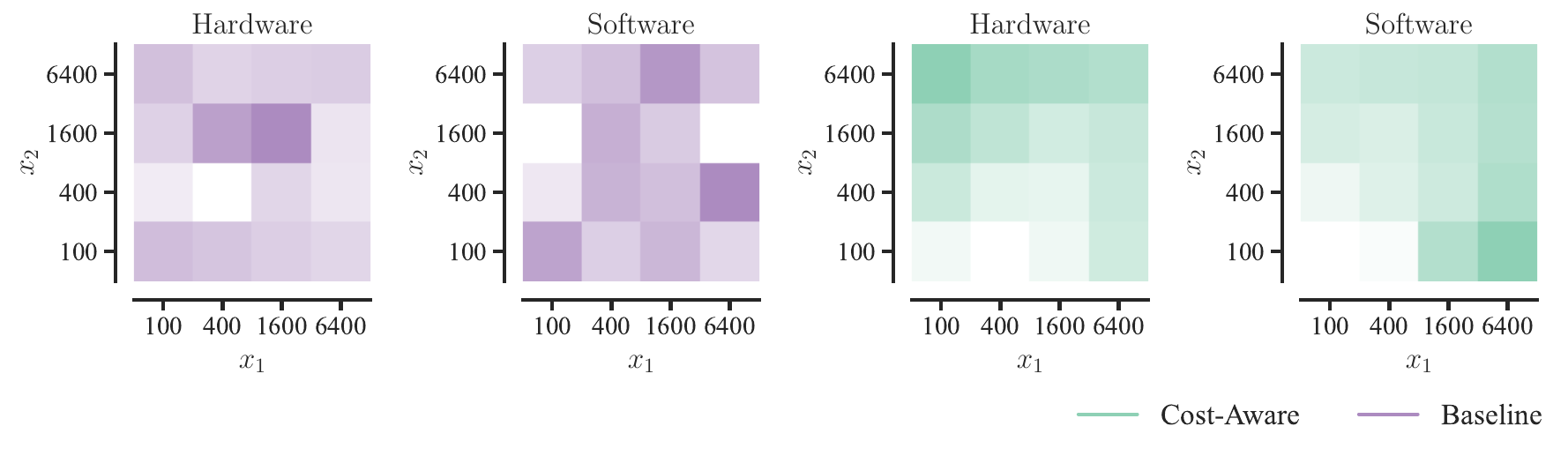}
        \subcaption{Levy %
        }
        \label{fig:levy_study3}
    \end{subfigure}

    \caption{\textbf{Study 3}:  Sample counts by method, component, and cost asymmetry. Heatmaps show the frequency of hardware and software edits under varying acquisition costs. The cost-aware method adjusts its sampling to avoid expensive actions, while the baseline distributes effort uniformly. Each heatmap is independently normalized for visibility.}
    \label{fig:study3_combined}
    \Description{Sampling patterns across hardware and software settings for three benchmark functions.
The figure shows heatmaps for Ackley, Goldstein–Price, and Levy.
For each function, baseline sampling (left heatmaps) emphasizes higher-cost regions in both hardware and software settings. The cost-aware method (right heatmaps) shifts sampling toward lower-cost regions, producing lighter patterns across the parameter grid. The specific sampling distributions vary by function but consistently show reduced emphasis on high-cost areas under cost-aware optimization.
}
\end{figure*}

\subsection{Study 3: Sensitivity to Cost Asymmetry}
This study evaluates whether the cost-aware method adapts to such cost asymmetries, whereas the baseline treats all evaluations uniformly. We independently varied hardware and software \create costs in \{100, 400, 1600, 6400\}, yielding 16 cost conditions. For each condition, we ran 50 optimization trials, each for 25 iterations. We recorded the number of samples taken by component and action type (\tweak, \swap, \create). A generalized linear mixed-effects model assessed whether the methods adjusted sampling behavior based on cost asymmetries.

\subsubsection*{Ackley.}
The cost-aware method adjusted sampling behavior to cost asymmetries (\Fig{fig:ackley_study3}). As hardware costs increased, it significantly reduced \create actions and shifted effort toward cheaper \swap and \tweak actions ($p < .001$ for interaction effects). The baseline showed no comparable adaptation, with hardware and software actions sampled at nearly constant rates. Post-hoc contrasts confirmed that the baseline systematically over-sampled hardware \create and under-sampled software edits ($p < .0001$). In contrast, the cost-aware method redistributed effort across action types in response to changing hardware costs, demonstrating that it adapts dynamically to asymmetries while the baseline remains static.

\subsubsection*{Goldstein–Price.}
The cost-aware method adapted strongly to cost asymmetries (\Fig{fig:goldstein_study3}). As hardware costs increased, it reduced \create sampling and shifted toward \swap and \tweak actions ($p < .001$ for interaction effects). In contrast, the baseline showed little adjustment, continuing to allocate similar effort across actions. Post-hoc contrasts confirmed that the baseline over-sampled hardware \create while under-sampling tweaks ($p < .0001$), whereas the cost-aware method redistributed sampling across modalities and action types in response to cost. These results indicate that only the cost-aware method responds dynamically to asymmetries, while the baseline remains rigid.

\subsubsection*{Levy.}
The cost-aware method again adapted to cost asymmetries (\Fig{fig:levy_study3}). With higher hardware costs, it reduced reliance on \create actions and increased use of \swap and \tweak ($p < .001$ for multiple contrasts). The baseline showed minimal adjustment, maintaining similar rates of hardware and software sampling. Post-hoc tests confirmed that the baseline consistently over-sampled \create relative to cheaper actions, while the cost-aware method redistributed effort more flexibly (e.g., significant contrasts between \create and both \swap and \tweak, all $p < .001$). These results reinforce that only the cost-aware method shifts sampling strategies in response to cost differences, whereas the baseline remains static.

\subsection{Study 5: Reweighting the Costs}
This study evaluates whether the cost-aware method scales effectively with increasing complexity and reallocates effort when costs change mid-cycle.  We evaluate two conditions over 24 iterations with three random initializations. In the \textit{constant} condition, cost weights remain fixed. In the \textit{dynamic} condition, hardware \create cost increases tenfold at iteration 10 (e.g., tool unavailability), and then drops to one-tenth at iteration 17 (e.g., faster fabrication found). We compare \create action frequency across three cost phases: iterations 4–10, 11–17, and 18–24. Each condition was repeated for 50 optimization trials. A generalized linear mixed-effects model assessed the interaction between condition, phase, and action type, focusing on \create sampling.

\subsubsection*{Ackley.}
The cost-aware method adapted strongly to cost changes. In the dynamic condition, \create sampling dropped sharply after the cost increase ($p < .0001$) and rose significantly when cost decreased again ($p < .0001$). By contrast, the baseline in the constant condition showed no meaningful shifts across phases (all $p > .47$). Between conditions, the dynamic run sampled fewer \create actions during the high-cost phase ($p < .0001$) and more after costs dropped ($p < .0001$), with no differences in the initial phase ($p = .76$). These effects occurred despite identical utility surfaces, demonstrating that the cost-aware method responds to cost structure rather than outcomes. These results show that the cost-aware method reallocates sampling effort in response to shifting constraints, while the baseline does not.

\subsubsection*{Goldstein–Price.}
The cost-aware method again adapted to cost changes. In the dynamic condition, \create sampling decreased slightly after the cost increase ($p = .16$) and rose strongly when cost dropped ($p < .0001$). The baseline in the constant condition showed no significant changes across phases (all $p > .59$). Between conditions, there were no differences in the initial phase ($p = .88$), but the dynamic run sampled fewer \create actions during the high-cost phase ($p = .008$) and more after costs dropped ($p < .0001$). These findings confirm that, even under a different ground truth, the cost-aware method reallocates sampling effort in line with shifting cost constraints, whereas the baseline remains static.

\subsubsection*{Levy.}
The cost-aware method again showed strong adaptation to shifting costs. In the dynamic condition, \create sampling dropped significantly after the cost increase ($p < .0001$) and rose sharply when costs decreased ($p < .0001$). The baseline in the constant condition showed no changes across phases (all $p \geq .74$). Between conditions, there was no difference in the initial phase ($p = .22$), but the dynamic run sampled fewer \create actions during the high-cost phase ($p < .0001$) and more when costs dropped again ($p < .0001$). These findings reinforce that the cost-aware method reallocates sampling effort in response to dynamic cost shifts, while the baseline remains static.

\subsection{Study 6: Effect of Inaccurate Cost Estimate}

\begin{figure*}[ht!]
    \centering

    \begin{subfigure}{0.49\linewidth}
        \centering
        \includegraphics[width=\linewidth]{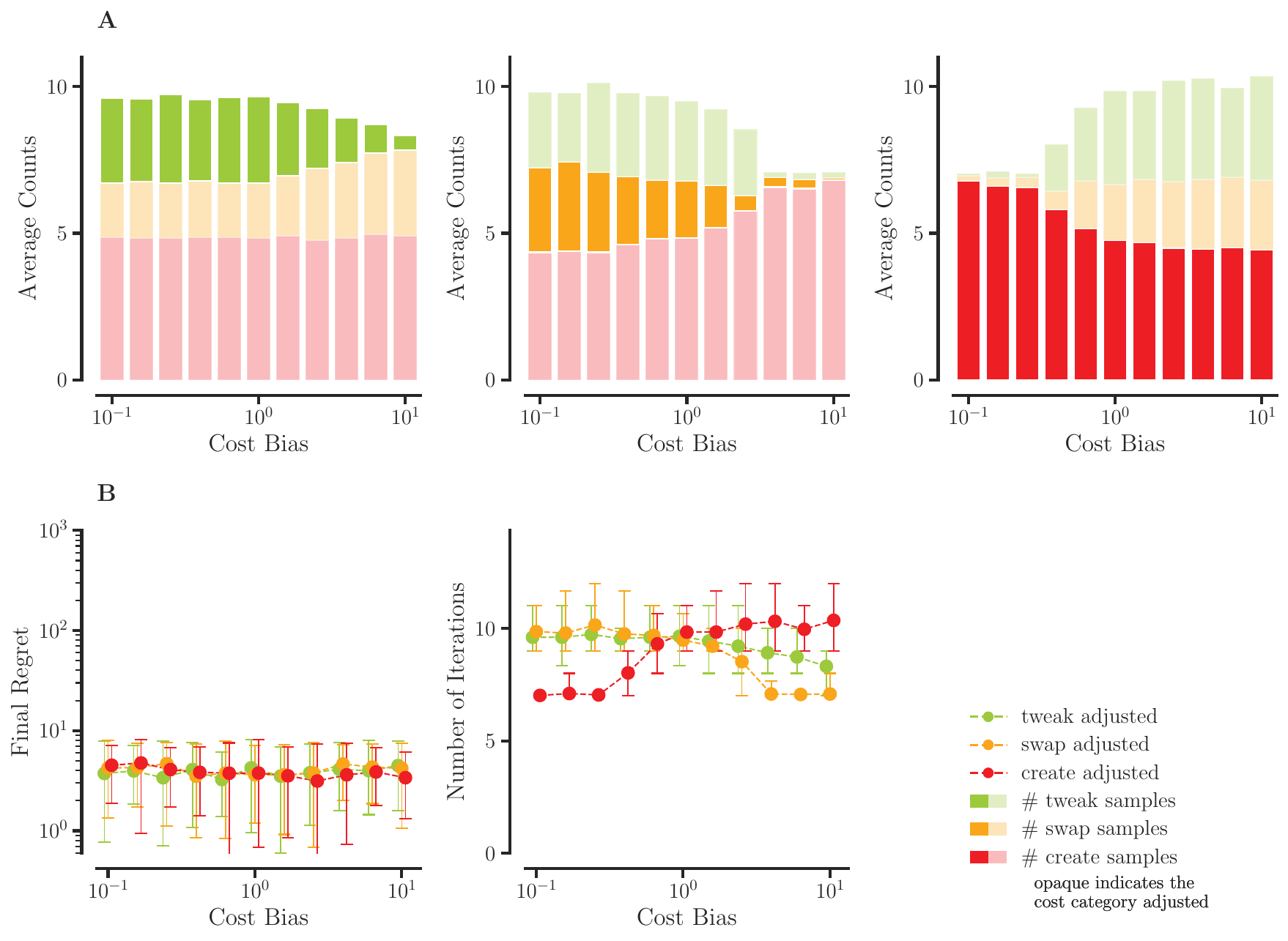}
        \subcaption{Ackley}
        \label{fig:ackley_study6}
    \end{subfigure}
    \hfill
    \begin{subfigure}{0.49\linewidth}
        \centering
        \includegraphics[width=\linewidth]{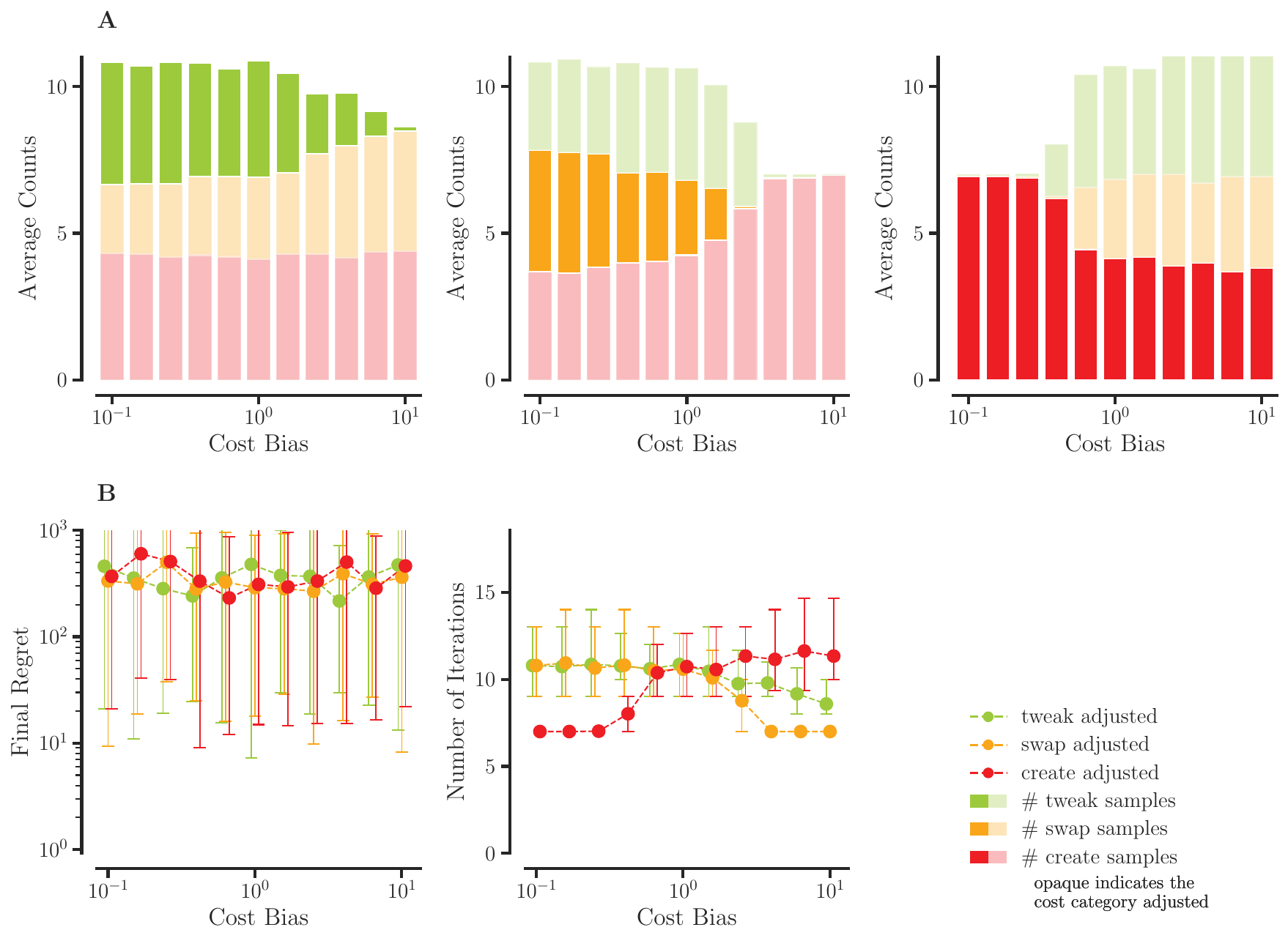}
        \subcaption{Goldstein--Price}
        \label{fig:goldstein_study6}
    \end{subfigure}

    \vspace{1em}

    \begin{minipage}[t]{0.49\linewidth}
        \centering
        \includegraphics[width=\linewidth]{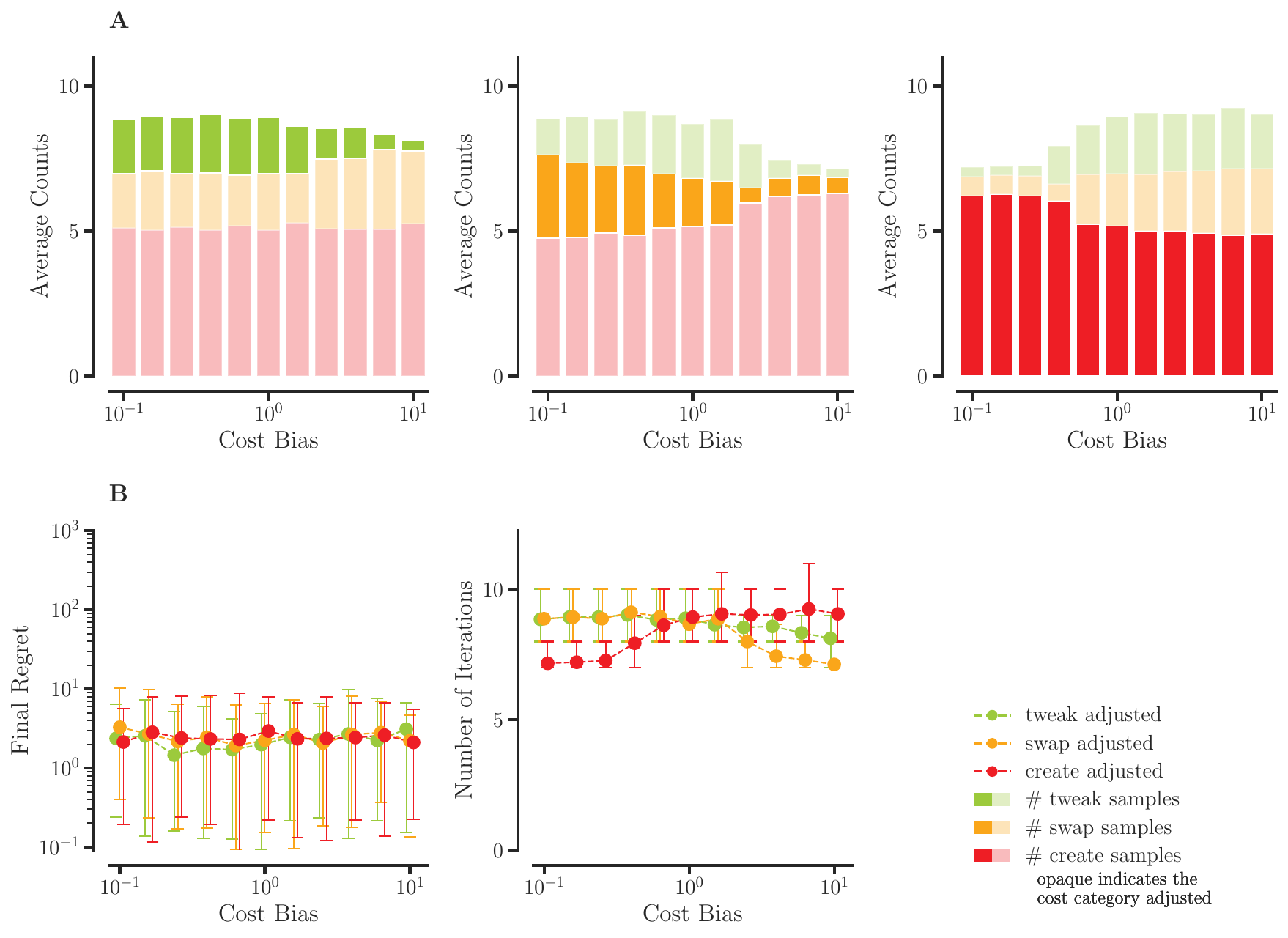}
        \subcaption{Levy}
        \label{fig:levy_study6}
    \end{minipage}
    \hfill
    \begin{minipage}[t]{0.49\linewidth}
        \vspace*{-1.7cm}
        \captionof{figure}{\add{\textbf{Study 6:} Effect of inaccurate cost estimates on optimization behavior. 
        The believed cost is a scaled version of the true cost by a factor $\alpha$ ranging from $10^{-1}$ to $10^{1}$. 
        The plots show how these misestimates affect final regret, iteration count, and sampling patterns across 
        \tweak, \swap, and \create. Error bars indicate 95\% confidence intervals.}}
        \Description{Effect of cost bias on regret, iteration count, and sampling behavior across benchmark functions.
        The figure presents results for Ackley, Goldstein–Price, and Levy.
        For each function, the top row shows two plots: final regret across bias values and the number of iterations taken. Regret remains fairly stable across bias settings, and iteration counts vary slightly depending on which action type is cost-adjusted.
        The bottom row shows three bar charts illustrating average sampling counts when tweak, swap, or create actions are cost-adjusted. Increasing bias shifts sampling away from the action whose cost is inflated and redistributes sampling toward lower-cost actions. The specific patterns differ across functions but exhibit the same overall trend.
        }
        \label{fig:app_study6}
    \end{minipage}

\end{figure*}

\paragraph{All} For the additional functions, we repeat the cost–misestimation experiment by scaling the true cost with $\alpha \in [10^{-1}, 10^{1}]$ and keeping the believed cost fixed throughout optimization. For all functions the qualitative behavior matches the main results. Increasing the believed cost of an operator reduces its sampling frequency, and decreasing it increases usage. The downstream effects follow the same structure: misestimation of \swap produces the strongest shifts, underestimation of \create quickly exhausts the budget, and misestimation of \tweak has comparatively mild influence. Despite these changes in sampling patterns, overall performance remains stable, and the method adapts predictably. As before, maintaining proportional cost relationships is more important than exact values.

\subsection{Summary}

\begin{table*}[t]
    \centering
    \begin{tabular}{p{4.5cm}cccc}
        \toprule
        \textbf{Experiment} & \textbf{Rosenbrock} & \textbf{Ackley} & \textbf{Goldstein-Price} & \textbf{Levy} \\
        \midrule
        Cost efficiency & \cmark & \cmark & \cmark & \cmark \\
        Budget constraints & \cmark & \xmark & \cmark & \cmark \\
        Cost asymmetries & \cmark & \cmark & \cmark & \cmark \\
        Changing cost mid-cycle & \cmark & \cmark & \cmark & \cmark \\
        Sensitivity to Cost Estimates & \cmark & \cmark & \cmark & \cmark \\
        \bottomrule
    \end{tabular}
    \caption{Performance of the cost-aware method across four experiments (rows) and four different ground truths (columns). Each cell indicates whether the capability was demonstrated under the given ground truth. The results for the Rosenbrock evaluation can be found in the main paper.}
    \label{tab:exp_gt}
    \Description{Comparison table with four experiments evaluated under four different ground truths. The table shows whether the cost-aware method demonstrated each capability under each condition.}
\end{table*}

Table~\ref{tab:exp_gt} summarizes results across experiments and ground truths. The cost-aware method consistently demonstrated cost efficiency (Exp.~1), effective adaptation to asymmetries (Exp.~3), and responsiveness to changing costs mid-cycle (Exp.~5) in all benchmarks. Under budget constraints (Exp.~2), its advantage was more mixed: it matched or outperformed the baseline on the Goldstein–Price and Levy functions, showed no reliable benefit on Ackley, and achieved consistent improvements on Rosenbrock. Overall, the findings confirm that cost-aware optimization generalizes across diverse functions, though its advantage under strict budget limits depends on problem characteristics.

}
\end{document}